\documentclass{emulateapj}
\pdfoutput=1
\usepackage{natbib}
\newcommand{\xmmn}{\textit{XMM-Newton}}
\newcommand{\task}[1]{\sc #1} 
\newcommand{\cxo}{\textit{Chandra X-ray Observatory}}
\newcommand{\chandra}{\textit{Chandra}}
\newcommand{\xmm}{\textit{XMM}}
\newcommand{\kepler}{\textit{Kepler}}
\newcommand{\sherpa}{\textit{Sherpa}}
\newcommand{\ee}[2]{\ensuremath{{#1}\times10^{#2}}}
\newcommand{\pcmsq}{\mbox{cm$^{-2}$}}

\shorttitle{X-RAY SOURCES IN NGC 6819}
\shortauthors{GOSNELL ET AL.}

\slugcomment{Submitted to ApJ 2011 June 29; accepted 2011 November 20}

\begin{document}
\title{An Unexpected Discovery in the Rich Open Cluster NGC 6819 Using XMM-Newton}

\author{Natalie M. Gosnell\altaffilmark{1}, David Pooley\altaffilmark{2,3}, Aaron M. Geller\altaffilmark{4}, Jason Kalirai\altaffilmark{5,6}, Robert D. Mathieu\altaffilmark{1}, Peter Frinchaboy\altaffilmark{7}, Enrico Ramirez-Ruiz\altaffilmark{8}}
\email{gosnell@astro.wisc.edu}
\altaffiltext{1}{Department of Astronomy, University of Wisconsin - Madison, 475 N. Charter St., Madison, WI  53706}
\altaffiltext{2}{Eureka Scientific, Inc., 2452 Delmer Street Suite 100, Oakland, CA  94602}
\altaffiltext{3}{Department of Astronomy, The University of Texas, Austin, TX, 78712}
\altaffiltext{4}{Center for Interdisciplinary Exploration and Research in Astrophysics (CIERA) and Department of Physics and Astronomy, Northwestern University, 2145 Sheridan Rd, Evanston, IL  60208}
\altaffiltext{5}{Space Telescope Science Institute, 3700 San Martin Drive, Baltimore, MD  21218}
\altaffiltext{6}{Center for Astrophysical Sciences, Johns Hopkins University, Baltimore, MD  21218}
\altaffiltext{7}{Texas Christian University, 2800 S. University Drive, Forth Worth, TX  76129}
\altaffiltext{8}{Department of Astronomy and Astrophysics, University of California, Santa Cruz, CA  95064}

\begin{abstract}
We present the first study of the X-ray population of the intermediate-age rich open cluster NGC 6819 using the \xmmn\ \textit{Observatory}. In the past decade, \textit{Chandra} X-ray observations have shown a relationship between the X-ray population of globular clusters and their internal dynamics and encounter frequency.  We investigate the role dynamics possibly play in the formation of X-ray sources in NGC 6819, and compare our results with known properties of field and globular cluster X-ray populations.  We implement a multi wavelength approach to studying the X-ray sources, utilizing X-ray and UV data from \xmm\ observations along with the wealth of photometry and radial-velocity data from the WIYN Open Cluster Study (WOCS) and the CFHT Open Cluster Survey.  Within the cluster half-light radius we detect 12 X-ray sources down to a luminosity of $10^{30}$  erg s$^{-1}$ for cluster members. The sources include a candidate quiescent low-mass X-ray binary (qLMXB), a candidate cataclysmic variable, and two active binary systems.  The presence of a qLMXB in an open cluster is previously unexpected given the known relationships between luminous X-ray sources and encounter frequency in globular clusters, and most likely has a dynamical origin.  
\end{abstract}

\keywords{binaries: close --- open clusters and associations: individual (NGC 6819) --- X-rays: binaries}

\section{INTRODUCTION}
It is well established that dynamical interactions play a large role in the formation of exotic binaries in globular clusters.  From the early days of X-ray astronomy, the overabundance of low-mass X-ray binaries (LMXBs) in outburst in globular clusters \citep{Clark75, Katz75} pointed to formation scenarios unique to the dense environments within globular clusters.  \citet{Verbunt87} showed that the number of luminous LMXBs known at the time was in agreement with formation theories based on close stellar encounters.  After the launch of the \cxo, a large number of quiescent LMXBs were found in globular clusters, and several groups independently established that these quiescent LMXBs, like those in outburst, were formed through dynamical interactions \citep{Gendre03, Heinke03, Pooley03}.

Cataclysmic variables (CVs) have also been predicted to have an enhancement in their formation rates in globular clusters \citep{Verbunt87, DiStefano94, Ivanova06}, although this was difficult to verify until \chandra\ began finding CVs in large numbers in globular clusters.  \citet{Pooley06} established that dynamics do play a role in the observed population of globular cluster CVs and that some component of the CV population is primordial in origin.  The relative numbers of the dynamically formed and primordial populations is subject to a number of poorly understood but important dynamical processes. For example, dense clusters would be expected to form many CVs through dynamical encounters.  However, similar encounters would be expected to destroy the wide progenitors of CVs before they entered the CV phase \citep{Davies97}.

Studies of low-density globular clusters \citep{Kong06, Bassa08, Lu09, Lan10} have revealed evidence that the X-ray source population (mainly CVs and chromospherically active main-sequence binaries) in these clusters is likely largely primordial in origin with few, if any, dynamically formed sources.  An extrapolation of this trend further down in mass and density to the realm of rich open clusters would predict almost zero X-ray sources, although significant populations are found in M67 \citep{vandenBerg04} and NGC 188 \citep{Gondoin05}.

Dynamics may be playing a larger role in open cluster environments than globular cluster studies would predict as well.  The high binary fraction of blue stragglers in NGC 188 and their unusual orbit parameters \citep{Mathieu09} could be indicative that dynamical encounters played a significant role in their production \citep{Leigh11}.  Complete $N$-body modeling of the open cluster M67 also points to dynamics playing a significant role in the formation of blue straggler populations \citep{Hurley05}.  Further, one of the binaries in NGC 6819 may be the result of an exchange encounter \citep{Gosnell07, Hole09}.

This evidence seems to indicate that dynamical interactions are playing a role in the production of exotic objects in rich open clusters, but the nature of that role is only beginning to be uncovered.  In order to gain a broader perspective on this, we have undertaken a systematic survey with \xmmn\ of eight rich open clusters previously unobserved in X-rays to be combined with archival analyses of four other clusters.  Our immediate aim is to identify and classify the X-ray sources through their X-ray, UV, and optical properties, and our ultimate goal is to investigate the role of dynamical interactions in rich open clusters in the production of luminous X-ray sources, which are usually close binaries.

We present here our results on the first cluster of our study, NGC 6819.  Physical parameters for this cluster have been estimated by numerous authors \citep[see][and references therein]{Kalirai01b, Hole09}.  Here, we adopt an extinction of $E(B-V) = 0.15$ given by \citet{Bragaglia01} and convert this to an equivalent column of $N_H = \ee{8.3}{20}~\pcmsq$ \citep{Predehl95}.  Recent \kepler\ data were used to determine a well-constrained distance of $2.34\pm0.06$ kpc and an age between 2 and 2.4 Gyr \citep{Basu11}.  We take the optical center of the cluster to be $19^{\rm h}41^{\rm m}18^{\rm s}$, $+40^{\circ}11'47''$ (J2000) \citep{Hole09}.

Our X-ray, UV, and optical observations are presented in \S\ref{sec:data}.  In \S\ref{sec:id}, we bring together the multi-wavelength data on each X-ray source within the half-light radius of NGC 6819.  Our X-ray flux estimates are given in \S\ref{sec:xrayspec}, and we discuss source classification in \S\ref{sec:classification}.  Our discussion is in \S\ref{sec:discussion} and our conclusions are laid out in \S\ref{sec:summary}.

\section{OBSERVATIONS AND DATA REDUCTION}
\label{sec:data}
We aim to investigate the X-ray sources of NGC 6819 as thoroughly as possible, and we therefore implement a multi wavelength study of X-ray source characteristics.  In addition to the \xmm\ X-ray data outlined in \S\ref{sec:xray}, we also include ultraviolet (UV) and optical photometry and spectroscopy.  We obtained UV photometry using the Optical Monitor aboard \xmm\ , the data and reduction of which are outlined in \S\ref{sec:uv}.  We also make use of the extensive optical photometry provided by the CFHT study of NGC 6819 \citep{Kalirai01b} and the photometry and radial-velocity information from the WIYN Open Cluster Survey\footnote{This is WIYN Open Cluster Study paper XLVII.} \citep[WOCS,][]{Mathieu00}, as outlined in \S\ref{sec:opt}.

\subsection{X-Ray Data}
\label{sec:xray}
\begin{figure}
\begin{center}
 \includegraphics[scale=0.45]{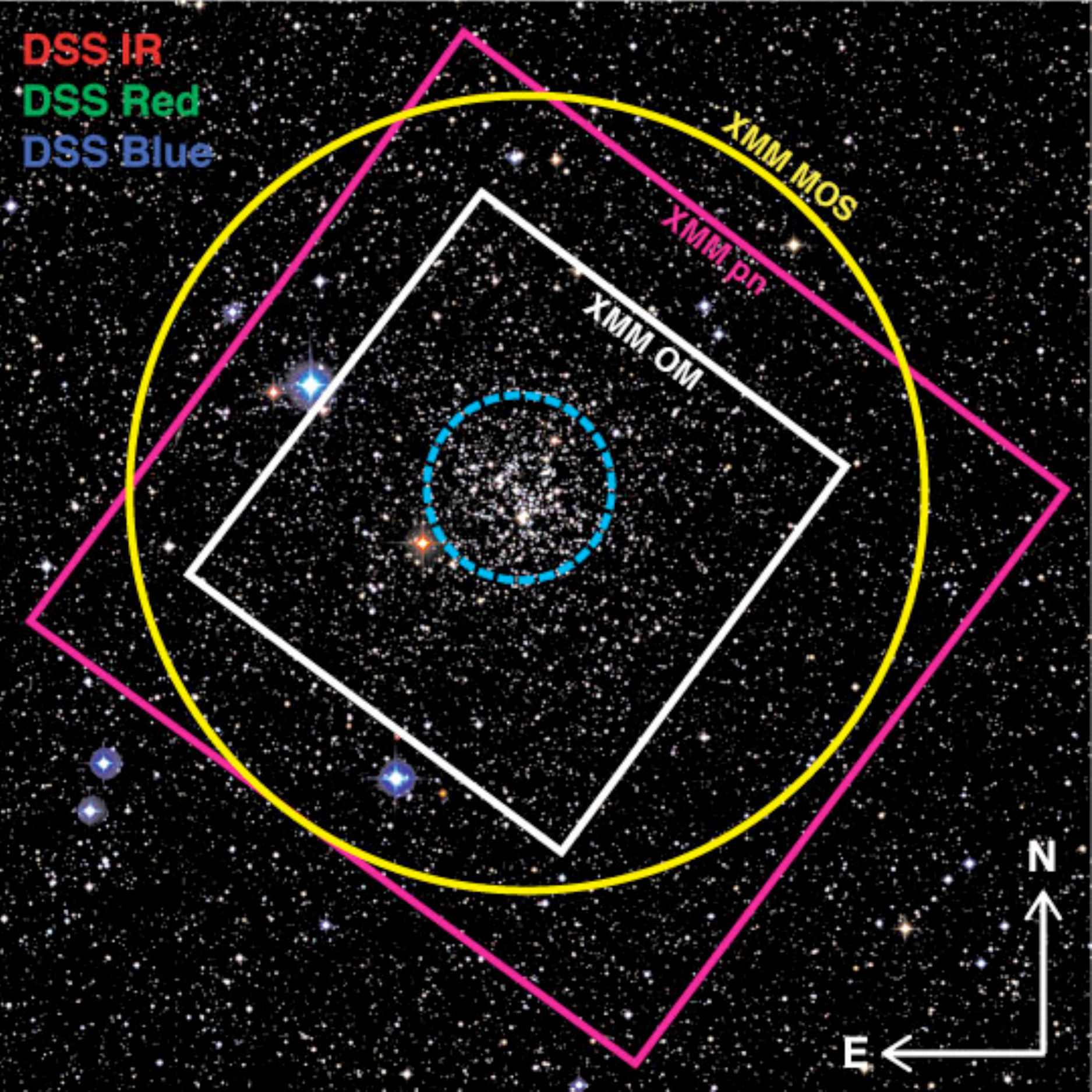}
 \end{center}
 \caption{RGB Digitized Sky Survey image of NGC 6819 overlaid with the \xmm\ field of view.  The different camera fields of view are indicated in the figure.  The 3\farcm3 half-light radius of NGC 6819 is shown with the cyan dashed circle.  The image is 39\arcmin\ on a side.}
\label{fov}
\end{figure}

NGC 6819 was observed by the \xmmn\ orbiting observatory for 25.0 ks (15:35:07--22:32:22 UT) on 2008 May 18 as part of our X-ray survey of rich open clusters.  The European Photon Imaging Camera (EPIC) pn camera and the two MOS cameras were used in full frame mode with the medium aluminum filter to block optical light.  The telescope boresight location was $19^{\rm h}41^{\rm m}18^{\rm s}$, $+40^{\circ}11'12''$ (J2000).  The field covered by \xmm\ is shown in Figure~\ref{fov}.  

We use \xmm\ Science Analysis Software (SAS) version 10.0 for our data analysis, following the routines outlined in the threads on the SAS help page\footnote{see http://xmm.esa.int/sas/current/documentation/threads/}.  Data from all three cameras are screened for high-energy background flaring, although we do not find flaring during the observation.  Filtered event lists are created using the {\task evselect} task to include only those events with energies between 0.2 keV and 10 keV.  We restrict the filtered events using PATTERN $<=12$ and $<=4$ for the MOS and pn cameras respectively, only allowing pixel patterns related to valid X-ray events that are well calibrated.  Images for each camera are created from the filtered event lists using {\task evselect}.  Individual tasks within the {\task edetect\_chain} task are run manually to detect sources.  The {\task emldetect} task uses filtered event lists from all three cameras simultaneously, resulting in a combined source list with position and photon counts.  This combined list contains 69 X-ray sources in our field.  We refer the reader to the SAS threads for a more detailed description of these routines.

\subsection{Ultraviolet Photometry}
\label{sec:uv}
UV data were obtained using the \xmm\ Optical Monitor (OM), a 30 cm telescope mounted next to the X-ray mirrors to allow simultaneous X-ray and UV or optical observations of the central $17' \times 17'$ X-ray field of view.  NGC 6819 was observed with the \textit{UVW1} filter (220--400 nm) for 5 ks and with the  \textit{UVM2} filter (200--280 nm) for two exposures of 5 ks each.  All data were taken in imaging mode.  The \textit{UVM2} images are aligned and stacked using the IRAF routine {\task imcombine} in order to reach a fainter limiting magnitude.  We implement the SAS pipeline routine {\task omichain} for the OM data reduction.  Within {\task omichain}, standard aperture photometry is applied to the detected sources, taking into account the point spread function of the instrument.  The output of {\task omichain} includes a combined source list for 3223 sources with calibrated data, including instrumental magnitudes, positions, and errors.  We use the interactive photometry routine {\task omsource} to determine UV magnitudes or limits for all X-ray sources, supplementing the pipeline results for coincident UV objects that are not initially detected.  

\subsection{Optical Photometry and Spectroscopy}
\label{sec:opt}
To identify optical counterparts, we utilize optical data from the the CFHT Open Cluster Survey \citep{Kalirai01a} and the WIYN Open Cluster Survey \citep[WOCS,][]{Mathieu00}.  Each of these surveys have obtained deep, multiband, wide-field imaging of the stellar populations of NGC 6819.  

For the CFHT Survey, \citet{Kalirai01b} used the CFH12K instrument to obtain multiple deep and shallow observations in two filters ($B$ and $V$) over a $42' \times 28'$ field of view that extends well beyond the cluster radius.  The multiple images are stacked together and PSF photometry is performed to measure the position and brightness of each source.  The final color-magnitude diagram (CMD) reveals a very tight main-sequence with several thousand member stars and extends down to $V$ = 24 (corresponding to a few tenths of a Solar mass).  The observations also reveal a rich white dwarf cooling sequence in the faint blue part of the CMD which stretches over four magnitudes \citep[Figure 7]{Kalirai01b}.  Further discussion of the data analysis of this cluster, and the derived parameters, are described in \citet{Kalirai01b} and \citet{Kalirai04}.

WOCS has over 10 years of high-precision radial-velocity (RV) measurements over a $23'$ square field of view centered on NGC 6819.  The WOCS primary target sample includes main-sequence, subgiant, giant, and blue straggler stars, spanning a magnitude range of $11 \leq V \leq 16.5$ with an approximate mass range of 1.1--1.6 M$_{\odot}$.  Using the Hydra MOS fiber-fed spectrograph on the WIYN\footnote{The WIYN Observatory is a joint facility of the University of Wisconsin - Madison, Indiana University, Yale University, and the National Optical Astronomy Observatories.} 3.5m telescope, \citet{Hole09} collected RV measurements of $\sim$2700 objects, each with a precision of $0.4$ km s$^{-1}$, with observations still ongoing.  The primary WOCS RV sample with $V \leq 15.0$ mag is $89\%$ complete and the secondary sample down to $V = 16.5$ mag is $\sim$60\% complete, resulting in 480 RV-determined cluster members.  These data also include orbital solutions for velocity variable stars within the completed sample with periods of up to $\sim$1000 days, providing valuable information for possible binary optical counterparts.  A more thorough discussion of the WOCS data are described in \citet{Hole09}, although we have continued observing the cluster since then and include additional RV data here.  A detailed description of the analysis techniques is provided by \citet{Geller08}.

\section{IDENTIFYING MULTI-WAVELENGTH COUNTERPARTS AND CLUSTER MEMBERS}
\label{sec:id}

Twelve of the 69 X-ray sources in the field of NGC 6819 lie within the cluster half-light radius of 3.3 arcmin.  The half-light radius is calculated from the star counts in \citet{Kalirai01b}. We only expect five or six background sources within this region using the log $N  - \,$log $S$ relationship from \citet{Giacconi01}.  Position and modeling results (discussed in \S\ref{sec:xrayspec}) for sources within the diameter of NGC 6819 are listed in Table~\ref{srctable1} and information for sources outside the diameter are listed in Table~\ref{srctable2}.  Sources within the cluster half-light radius are numbered X1 through X12 according to their total net counts in the EPIC camera.  Source position errors are calculated using the method of the \textit{XMM} Serendipitous Survey by adding the statistical $1\sigma$ errors from {\task emldetect} to the accepted systematic \textit{XMM} position error of 1\farcs0 in quadrature \citep{Watson09}.

\begin{deluxetable*}{llccccccc}
\tabletypesize{\scriptsize}
\tablecolumns{9}
\tablewidth{0pt}
\tablecaption{\textit{XMM} X-ray sources within the half-light radius of NGC 6819\label{srctable1}}
\tablehead{
  \colhead{}					&
  \colhead{}					&
  \colhead{Error\,\tablenotemark{b} }	&
  \colhead{}					&
  \colhead{}					&
  \colhead{}					&
  \colhead{Reduced}				&
  \colhead{}					&
  \colhead{}					\\
  \colhead{Source\,\tablenotemark{a}} 		& 
  \colhead{Position (J2000)} 		& 
  \colhead{(arcsec)}				&
  \colhead{Counts\,\tablenotemark{c}} 		& 
  \colhead{L$_{X}$ (ergs s$^{-1}$)\,\tablenotemark{d}}	&
  \colhead{Model\,\tablenotemark{e}}		&
  \colhead{Statistic}				&
  \colhead{Q-value}				&
  \colhead{Comments}                          }
\startdata
X1   &  19 41 14.08, +40 14 05.60  &  1.07 &  816   & $(1.05\pm0.01)\times10^{32}$  & kT$_{\mathrm{NSA}}=62.8^{+0.5}_{-0.5}$ & 0.88 & 1.0 & qLMXB \\   	
X2   &  19 41 29.19, +40 11 23.83  &  1.07 &  727   & $(6.8\pm0.5)\times10^{31}$  & kT$_{\mathrm{APEC}}=6.2^{+2.1}_{-1.0}$ & 0.96 & 0.98 & CV \\  
X3   &  19 41 25.41, +40 11 44.63  &  1.16 &  361   & $(3.7\pm0.5)\times10^{31}$  & $\alpha$=1.5$^{+0.2}_{-0.8}$  &  0.70 & 1.0 & \\	
X4   &  19 41 13.30, +40 12 48.14  &  1.21 &  292   & $(2.5\pm0.3)\times10^{31}$  & kT$_{\mathrm{APEC}}=4.1^{+2.6}_{-1.1}$ &  0.76 & 1.0  & \\	
X5   &  19 41 22.47, +40 13 06.34  &  1.27 &  267   & $(2.1\pm0.4)\times10^{31}$  & kT$_{\mathrm{BB}}=0.9^{+0.1}_{-0.1}$ &  0.82 & 1.0  & \\	
X6   &  19 41 22.02, +40 12 01.18  &  1.40 &  155   & $(1.6\pm0.1)\times10^{31}$  & $\alpha$=3.0$^{+0.4}_{-0.3}$ &  0.66 & 1.0 & RS CVn\\	
X7   &  19 41 13.86, +40 09 32.70  &  1.55 &  114   & $(8.4\pm1.9)\times10^{30}$ & $\alpha$=2.2$^{+0.4}_{-0.3}$  &  1.03   &  1.0  &    \\   
X8   &  19 41 06.34, +40 11 42.42  &  1.79 &  113   & $(1.1\pm0.1)\times10^{31}$  & $\alpha$=2.7$^{+0.4}_{-0.4}$ &  0.56 & 1.0 & \\	
X9   &  19 41 25.89, +40 12 23.63  &  1.84 &  111   & $(9.3\pm1.3)\times10^{30}$  & $\alpha$=2.7$^{+0.4}_{-0.4}$ &  0.60 & 1.0 & Active Binary \\	
X10   &  19 41 11.81, +40 10 43.01  &  1.96 & \phn91  & $(9.1\pm4.9)\times10^{30}$ & $\alpha$=2.2$^{+0.8}_{-0.5}$ &  0.54 & 1.0 & \\	
X11  &  19 41 18.48, +40 14 15.95  &  1.89 & \phn88  & $(8.7\pm1.2)\times10^{30}$ & $\alpha$=2.8$^{+0.5}_{-0.4}$ &  0.60 & 1.0 & \\	
X12  &  19 41 26.64, +40 12 57.70  &  1.76 & \phn86  & $(1.4\pm0.5)\times10^{31}$ & $\alpha$=1.4$^{+0.3}_{-0.3}$ &  0.59 & 1.0 & \\	
\enddata
\tablenotetext{a}{Numbered according to their total counts in 0.2--10.0 keV band}
\tablenotetext{b}{Total 1$\sigma$ position error}
\tablenotetext{c}{Net EPIC camera counts in 0.2--10.0 keV band}
\tablenotetext{d}{Unabsorbed luminosity between 0.2--10.0 keV}
\tablenotetext{e}{Model parameter listed for sources X1-X5 based on best fit.  Parameters correspond to effective temperature for a neutron star atmosphere (kT$_{\mathrm{NSA}}$) in eV, APEC plasma temperature (kT$_{\mathrm{APEC}}$) in keV, thermal blackbody temperature (kT$_{\mathrm{BB}}$) in keV, or powerlaw index ($\alpha$).} 
\end{deluxetable*}

\begin{deluxetable*}{ccccc}
\tabletypesize{\scriptsize}
\tablecolumns{5}
\tablewidth{300pt}
\tablecaption{\textit{XMM} X-ray sources outside the half-light radius of NGC 6819\label{srctable2}}
\tablehead{
  \colhead{}						&
  \colhead{Position Err. }				&
  \colhead{}						&
  \colhead{F$_{X} (\times10^{-14}$) \tablenotemark{c}}	&
  \colhead{}						\\
  \colhead{Source} 					& 
  \colhead{(arcsec)\,\tablenotemark{a}}	&
  \colhead{Counts\,\tablenotemark{b}} 			& 
  \colhead{(ergs cm$^{2}$ s$^{-1}$ )}  &
  \colhead{PL Index\,\tablenotemark{d}}					
}
\startdata
XMMUJ 194157.6+401514  &   1.02 &   3167  & $21.4\pm0.5$ & $3.10\pm0.04$  \\  	
XMMUJ 194216.0+401046  &   1.04 &   1823  & $24.0\pm1.3$ & $2.07\pm0.05$ \\  	
XMMUJ 194046.7+402153  &   1.07 &   1227  & $13.8\pm0.5$ & $2.95\pm0.07$ \\   	
XMMUJ 194121.1+400044  &   1.15 & \phn504 & $3.5\pm0.2$ & $3.2\pm0.1$  \\  	
XMMUJ 194206.7+401441  &   1.15 & \phn472 & $5.1\pm0.8$ & $2.9\pm0.1$  \\  	
XMMUJ 194200.0+401520  &   1.14 & \phn458 & $4.1\pm0.3$ & $2.7\pm0.1$  \\  	
XMMUJ 194057.3+402031  &   1.21 & \phn397 & $3.7\pm0.6$ & $1.44\pm0.09$  \\ 	
XMMUJ 194049.0+400615  &   1.22 & \phn378 & $2.3\pm0.4$ & $1.8\pm0.1$  \\ 	
XMMUJ 194100.2+400220  &   1.20 & \phn336 & $5.1\pm0.7$ & $1.7\pm0.1$  \\  	
XMMUJ 194146.6+401014  &   1.10 & \phn334 & $5.7\pm0.7$ & $2.1\pm0.1$   \\   
XMMUJ 194050.1+401650  &   1.20 & \phn331 & $3.6\pm0.4$ & $2.1\pm0.1$   \\   
XMMUJ 194212.7+401747  &   1.13 & \phn330 & $9.8\pm1.3$ & $1.68\pm0.09$   \\   
XMMUJ 194030.8+400641  &   1.30 & \phn312 & $3.4\pm0.8$ & $3.2\pm0.2$   \\   
XMMUJ 194224.4+400929  &   1.37 & \phn309 & $3.5\pm1.4$ & $3.5\pm0.1$   \\   
XMMUJ 194104.8+400100  &   1.15 & \phn307 & $5.7\pm1.2$ & $1.2\pm0.1$   \\   
XMMUJ 194209.9+401449  &   1.25 & \phn295 & $4.8\pm0.9$ & $1.8\pm0.2$   \\   
XMMUJ 194141.5+400658  &   1.30 & \phn292 & $4.8\pm1.7$ & $0.7\pm0.1$   \\   
XMMUJ 194153.0+401145  &   1.17 & \phn274 & $5.7\pm1.2$ & $1.6\pm0.1$   \\   
XMMUJ 194053.4+401556  &   1.32 & \phn247 & $5.4\pm3.5$ & $2.9\pm0.5$   \\   
XMMUJ 194145.4+400806  &   1.48 & \phn245 & $1.5\pm0.6$ & $2.2\pm0.3$   \\   
XMMUJ 194137.5+400523  &   1.14 & \phn218 & $5.5\pm1.3$ & $1.2\pm0.1$   \\   
XMMUJ 194147.0+401347  &   1.34 & \phn203 & $3.0\pm1.0$ & $1.3\pm0.2$   \\   
XMMUJ 194115.8+401724  &   1.40 & \phn178 & $0.8\pm0.1$ & $3.3\pm0.2$   \\   
XMMUJ 194034.9+401152  &   1.60 & \phn178 & $2.6\pm2.8$ & $3.2\pm0.5$   \\   
XMMUJ 194153.5+401531  &   1.52 & \phn177 & $2.6\pm0.6$ & $1.9\pm0.2$   \\   
XMMUJ 194105.5+401430  &   1.42 & \phn163 & $1.5\pm0.6$ & $1.7\pm0.2$   \\   
XMMUJ 194131.3+402033  &   1.53 & \phn161 & $1.5\pm0.6$ & $1.6\pm0.3$   \\   
XMMUJ 194057.1+395959  &   1.36 & \phn157 & $0.8\pm0.1$ & $2.8\pm0.2$   \\   
XMMUJ 194106.6+400753  &   2.01 & \phn133 & $1.1\pm0.3$ & $2.4\pm0.3$   \\   
XMMUJ 194200.2+400516  &   1.89 & \phn132 & $0.5\pm0.1$ & $2.8\pm0.3$   \\   
XMMUJ 194139.5+402141  &   2.07 & \phn128 & $0.6\pm0.4$ & $1.9\pm0.4$   \\   
XMMUJ 194013.1+401143  &   1.88 & \phn128 & $1.3\pm0.7$ & $2.2\pm0.4$   \\   
XMMUJ 194226.2+400934  &   1.74 & \phn122 & $0.7\pm0.5$ & $2.7\pm0.3$   \\   
XMMUJ 194209.0+401113  &   1.70 & \phn122 & $1.6\pm0.7$ & $1.9\pm0.3$   \\   
XMMUJ 194037.0+401141  &   1.59 & \phn112 & $1.1\pm0.2$ & $3.3\pm0.2$   \\   
XMMUJ 194121.9+402336  &   2.93 & \phn111 & $0.9\pm0.2$ & $3.5\pm0.3$   \\   
XMMUJ 194041.5+400745  &   1.68 & \phn109 & $1.5\pm0.7$ & $1.6\pm0.3$   \\   
XMMUJ 194155.3+400940  &   1.49 & \phn107 & $1.2\pm0.3$ & $2.6\pm0.3$   \\   
XMMUJ 194149.9+400804  &   1.65 & \phn102 & $1.6\pm2.4$ & $3.1\pm0.5$   \\   
XMMUJ 194156.2+401351  &   1.92 & \phn101 & $2.5\pm2.0$ & $1.6\pm0.5$   \\   
XMMUJ 194056.2+401541  &   1.71 & \phn\phn99 & $2.6\pm2.8$ & $1.4\pm0.2$   \\   
XMMUJ 194040.2+401611  &   1.62 & \phn\phn98 & $1.2\pm0.3$ & $1.6\pm0.2$   \\   
XMMUJ 194037.7+400421  &   1.99 & \phn\phn96 & $9.2\pm3.8$ & $1.8\pm0.9$   \\   
XMMUJ 194055.4+401242  &   1.87 & \phn\phn92 & $0.7\pm0.1$ & $2.6\pm0.3$   \\   
XMMUJ 194202.4+400734  &   2.04 & \phn\phn89 & $0.8\pm0.8$ & $2.5\pm0.6$   \\   
XMMUJ 194141.1+401522  &   1.88 & \phn\phn86 & $1.6\pm1.2$ & $2.1\pm0.5$   \\   
XMMUJ 194113.0+400629  &   2.45 & \phn\phn85 & $5.2\pm2.4$ & $1.0\pm0.3$   \\   
XMMUJ 194054.2+401036  &   1.86 & \phn\phn81 & $1.6\pm1.4$ & $1.3\pm0.5$   \\   
XMMUJ 194021.4+401546  &   2.38 & \phn\phn80 & $0.3\pm0.2$ & $1.8\pm0.4$   \\   
XMMUJ 194218.7+401544  &   1.77 & \phn\phn78 & $1.6\pm0.7$ & $1.1\pm0.4$   \\   
XMMUJ 194143.4+401600  &   1.74 & \phn\phn73 & $0.3\pm0.2$ & $1.6\pm0.4$   \\   
XMMUJ 194130.0+400348  &   2.04 & \phn\phn66 & $0.6\pm0.2$ & $3.7\pm0.7$   \\   
XMMUJ 194052.9+400400  &   2.06 & \phn\phn66 & $0.6\pm0.2$ & $3.1\pm0.4$   \\   
XMMUJ 194205.7+402102  &   2.35 & \phn\phn60 & $3.9\pm1.5$ & $1.9\pm0.3$   \\   
XMMUJ 194200.3+402202  &   2.16 & \phn\phn53 & $0.9\pm0.2$ & $3.8\pm0.4$   \\   
XMMUJ 194200.0+401303  &   2.40 & \phn\phn53 & $0.7\pm0.4$ & $3.1\pm0.8$   \\   
XMMUJ 194101.7+400411  &   1.87 & \phn\phn52 & $0.7\pm0.4$ & $2.1\pm0.5$   \\   
\enddata
\tablenotetext{a}{1$\sigma$ error calculated by \task{emldetect}}
\tablenotetext{b}{Net EPIC camera counts in 0.2--10.0 keV band}
\tablenotetext{c}{Model-determined absorbed flux between 0.2--10.0 keV}
\tablenotetext{d}{Best fit power law index}
\end{deluxetable*}

\subsection{X-ray Source Cross-Correlation}

\subsubsection{Astrometry}
\begin{figure}
\begin{center}	
 \includegraphics[scale=0.65]{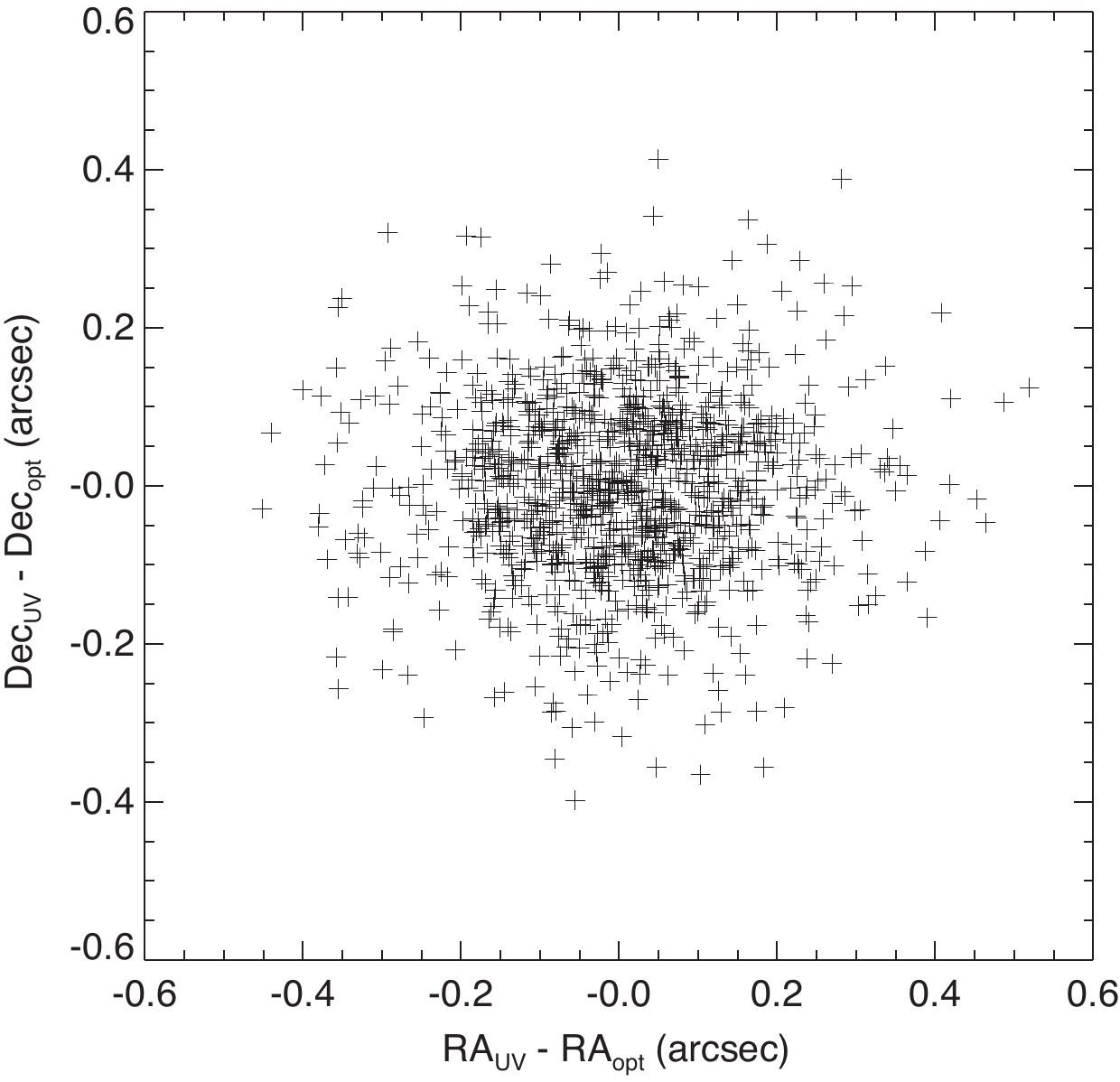}
\end{center}
\caption{Residuals of the corrected \xmm\ OM astrometry after matching \xmm\ UV sources to optical counterparts in the WOCS catalog.}
\label{astplot}
\end{figure}

Corrections for systematic offsets between \xmm\ and optical positions are necessary before completing successful source cross-correlation.  The WOCS NGC 6819 optical catalog has an absolute astrometric error of only 50 to 100 mas \citep{Hole09}, so we use its positions as a baseline.  We match the \xmm\ frame to the optical frame by correlating the OM and optical positions through an iterative sigma clipping technique.  First, we overlay WOCS object positions on the \textit{UVW1} OM image and identify 20 bright counterparts by eye.  From these sources we compute average offsets in right ascension and declination.  This offset is applied to all 3223 UV source positions and is cross-correlated to the optical catalog.  Sources are considered a match if they fall within the {\task omichain} $1\sigma$ UV position error.  Using the new counterpart list we calculate the average offset and apply an updated correction to the original UV source positions.  This process is continued until no new counterparts are found, resulting in a total of 1150 optical-UV counterparts.  The final average offsets are $-3\farcs75 \pm 0\farcs14$ and $0\farcs65 \pm 0\farcs11$ in right ascension and declination, respectively.  This average offset is applied to all UV and X-ray source positions.  The residuals of our astrometry correction have a sigma of 0\farcs17 and are shown in Figure~\ref{astplot}.

\subsubsection{Cross-Correlation}
\begin{figure*}
\begin{center}
 \includegraphics[scale=1.0]{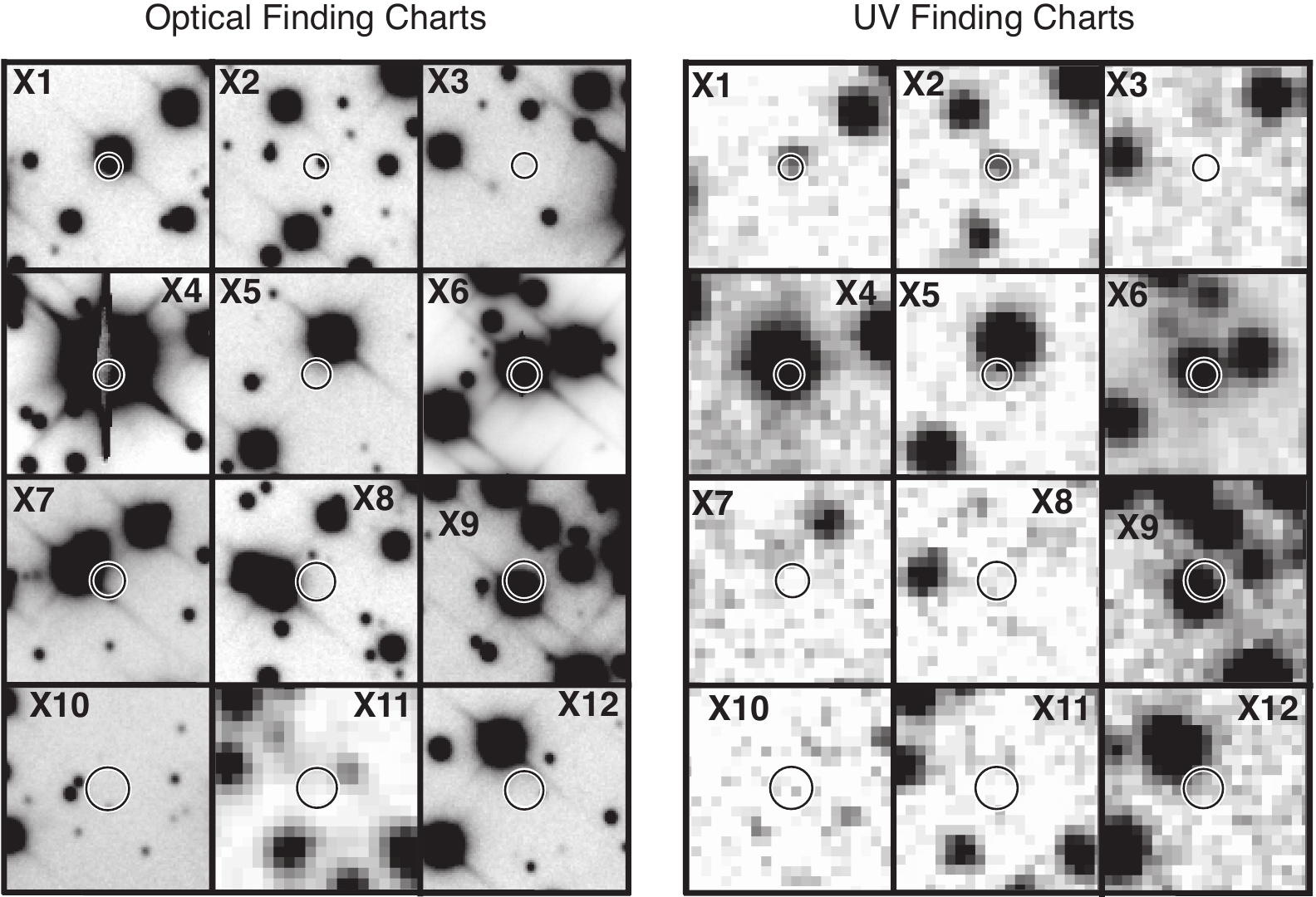}
\end{center}
\caption{$V$-band and \textit{UVW1} finding charts for X-ray sources within the half-light radius of NGC 6819.  The optical images are from CFHT observations of the cluster \citep{Kalirai01b}.  Each box is 20\farcs0 on a side overlaid with the corresponding X-ray position error circle.  The greyscale for the optical X6 chart is lighter for clarity.  The optical chart for source X10 is shown on a DSS image of NGC 6819 due to its location between chips in the CFHT images. Sources X7 and X9 are shown using the \textit{UVM2} image due to their position near dead pixels in the \textit{UVW1} image.}
\label{findingcharts}
\end{figure*}

We aim to classify the X-ray sources using their broadband spectral characteristics, which is heavily aided by identifying UV and optical counterparts.  After applying the astrometry corrections, the {\task emldetect} X-ray source list is correlated to the UV, CFHT, and WOCS optical source catalogs within the X-ray position error.  Optical and UV finding charts for sources within the half-light radius of the cluster are shown in Figure~\ref{findingcharts}.

\subsection{Identifying Members}
The UV and optical properties of X-ray source counterparts are listed in Table~\ref{match}.  Sources without optical or UV counterparts are omitted from this table.  An optical or UV object is considered a probable counterpart if it is separated by less than $1\sigma$ from the X-ray source.  Optical or UV sources with a separation between $1\sigma$ and $1.5\sigma$ are considered possible counterparts; their information is shown in the table with italics.  We limit our possible counterparts to within 1.5$\sigma$ in order to reduce the number of possible spurious counterpart identifications that may come from the relatively large X-ray position errors.  While some X-ray sources associated with the cluster may be located outside the half-light radius, their frequency becomes much smaller than that of unrelated background sources.  For example, out to twice the half-light radius (6.6 arcmin), we detect 28 X-ray sources, and we expect between 24 and 32 background sources using the log $N$ - log $S$ relationship of \citet{Giacconi01}.  For this reason, we restrict our analysis to only those sources within the half-light radius.

\begin{deluxetable*}{clccccccccc}
\tabletypesize{\scriptsize}
\tablecolumns{10}
\tablewidth{0pt}
\tablecaption{X-ray Source Optical and UV Counterpart Properties\label{match}}
\tablehead{
  \colhead{X-ray ID} 			& 
  \colhead{WOCS ID\tablenotemark{a}}	&
  \colhead{$V$\tablenotemark{b}}	&
  \colhead{$B-V$\tablenotemark{b}} 	& 
  \colhead{\textit{UVW1}} 		&
  \colhead{\textit{UVM2}} 		&
  \colhead{PM prob.\tablenotemark{c}}	&
  \colhead{RV prob. \tablenotemark{d}}	&
  \colhead{Opt. dis.\tablenotemark{e}}	&
  \colhead{UV dis.}			&
}
\startdata
\sidehead{Within half-light radius of cluster} 
X1		 &	\nodata 	&	\nodata  	 	&  \nodata  		&  \textit{19.75}\tablenotemark{f}  &  \textit{19.93}  	&  \nodata  	&  \nodata 	&  \nodata  	  &  \textit{1.45}  	\\
X2    		 &	\nodata		&	\textit{21.40}	 	&  \textit{0.50}	&	19.69              	    &   20.11  		&  \nodata 	&  \nodata 	&   \textit{1.24} &   0.79		\\
X4		 &	\textit{4003} 	&	\textit{13.06}   	&   \textit{1.41}   	&   \textit{15.43}  		    &  \textit{17.99}   &  \textit{0}  	&  \textit{2}  	&  \textit{1.57}  &   \textit{1.64} 	\\
X6    		 &	3002     	&	12.76\tablenotemark{a}  &  1.12    		& 	15.44   		    &   16.98  		&   	99  	&	 93 	& 	0.98      &  	1.16		\\
X9    		 &	52004		&	15.65   		&  0.844    		&  17.28   			    &   18.76  		&   	96 	& \nodata	&	1.32      &  	1.14		\\
\sidehead{Outside half-light radius of cluster}
194157.6+401514  &  	1017     &  	6.23    &	0.17    &  \nodata	 &	\nodata	 &	0	& \nodata	&	0.41    &   \nodata 	\\       
194216.0+401046  &	264023   & 	18.36   &	0.59    &  \nodata	 &	\nodata	 &	\nodata & \nodata	&	0.86    &   \nodata	\\       
194200.0+401520  & 	64018    &  	16.58   &	1.52    &  \nodata	 &	\nodata	 &	\nodata	& \nodata	&	0.61    &   \nodata	\\       
194121.1+400044  & 	\nodata	 &    \nodata	&	\nodata &  20.40   	 &  19.31  	 &     \nodata 	&  \nodata 	&    \nodata	& 	1.27	\\  
194206.7+401441  &	204020   &	18.14   &	1.55    &  \nodata	 &	\nodata	 &	\nodata	& \nodata	&	0.96    &   \nodata	\\      
194141.5+400658  & 	\nodata  & 	20.08   &	0.52    &  19.73   	 &  21.01  	 &     \nodata 	& \nodata	&	0.84    & 	0.90	\\
194115.8+401724  & 	3012     &  	13.21   &	0.19    &  13.90  	 &  15.52  	 &  	 0	& 0		&	1.08    & 	1.23	\\
194105.5+401430  & 	99008    & 	16.98   &	1.65    &  \nodata	 &	\nodata	 &	\nodata	& \nodata	&	1.25    &   \nodata	\\     
194106.6+400753  & 	\nodata  & 	22.44   &	1.42	&  \nodata	 &	\nodata	 &	\nodata	& \nodata	&	1.77    &   \nodata	\\        
194041.5+400745  & 	234016   & 	18.69   &	0.38    &  19.81  	 &  19.69  	 &  \nodata 	& \nodata	&	1.44   	&	1.58	\\
194140.1+400029  &	169025   & 	 17.52  &	1.60    &  \nodata	 &	\nodata	 &	\nodata	& \nodata	&	1.98    &   \nodata	\\     
194113.0+400629  & 	\nodata  & 	 21.33  &	1.49	&  \nodata 	 &	\nodata	 &	\nodata	& \nodata	&	0.36    &   \nodata	\\    
194121.9+402336  & 	\nodata  &	17.56   &	0.76    &  \nodata	 &	\nodata	 &	\nodata	& \nodata	&	1.96    &   \nodata	\\	   
194141.1+401522  &	255012   &	19.29   &	0.79    &  \nodata	 &	\nodata	 &	\nodata	& \nodata	&	1.58    &   \nodata	\\      
194132.6+400009  & 	22024    &	14.48   &	0.63    &  \nodata	 &	\nodata	 &	0	& 0		&	1.75    &   \nodata	\\      
194052.9+400400  & 	\nodata	 &	21.73   &	1.58    &  \nodata	 &	\nodata	 &	\nodata	& \nodata	&	0.81    &   \nodata	\\	
\enddata
\tablenotetext{a}{\citet{Hole09}}
\tablenotetext{b}{Values from \citet{Kalirai01b}, unless otherwise noted}
\tablenotetext{c}{Proper motion determined cluster membership probability (I. Platais, private communication)}
\tablenotetext{d}{Radial-velocity determined cluster membership probability \citep{Hole09}}
\tablenotetext{e}{Distance between optical and UV counterparts and the X-ray source are given in arcseconds}
\tablenotetext{f}{Italicized information corresponds to possible counterparts separated from the X-ray source by more than the $1\sigma$ X-ray position error}
\end{deluxetable*}

\begin{figure}
\begin{center}
 \includegraphics[scale=0.45]{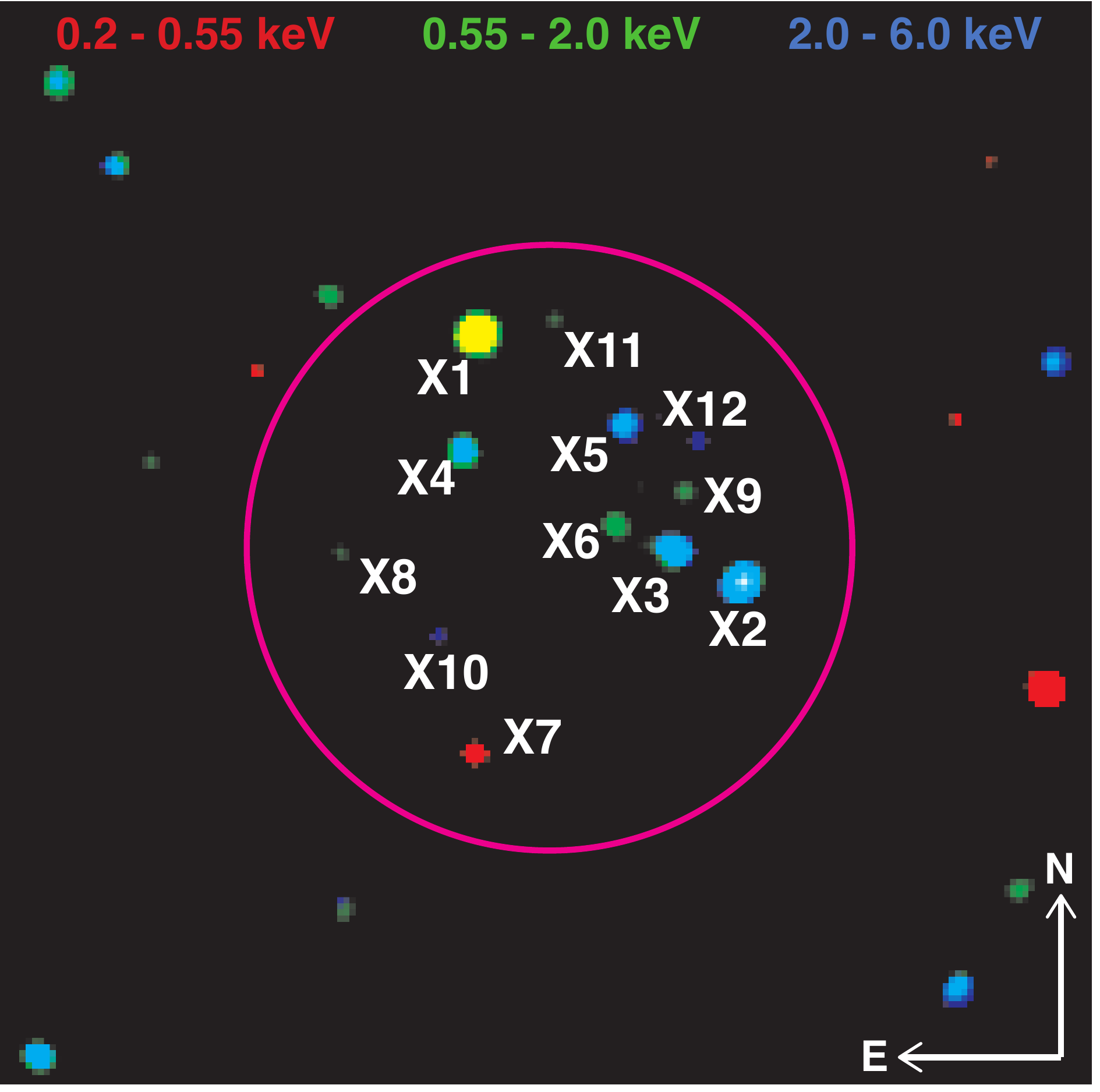}
\end{center}
\caption{X-ray ``RGB" image of the inner field of NGC 6819.  The red, green, and blue colors correspond to low, medium, and high X-ray energy ranges as indicated in the figure.  The magenta circle shows the 3\farcm3 half-light radius of the cluster.  The 12 sources within the half-light radius are numbered according to their total counts in the \xmm\ EPIC camera.  The image is 12\arcmin\ on a side.}
\label{xrayrgb}
\end{figure}

We also restrict possible X-ray source cluster membership to these sources.  Sources within the half-light radius have optical counterparts with high proper motion membership probability and also include the brightest sources in the field outside of obvious foreground stars.  Optical counterparts of sources outside the half-light radius for which RV and proper motion data are available are shown to be non-members, as can be seen in Table~\ref{match}.  For these reasons we consider it unlikely that we miss X-ray cluster members by restricting our sample to sources within the half-light radius and therefore focus on these objects in the following sections.  A ``true color'' X-ray image of this region is shown in Figure~\ref{xrayrgb}.  Sources within the half-light radius are numbered X1 to X12 according to their detected counts in 0.2--10.0 keV from all three EPIC cameras.

\section{X-RAY SPECTROSCOPY}
\label{sec:xrayspec}
In order to obtain fluxes and characterize the X-ray sources, we extract and fit spectra for all 69 sources, although more thorough model fitting is only carried out on sources X1-X12.  Spectral extraction is completed on the filtered event list for each camera using the {\task evselect} task.  Some sources fall between CCD chips or outside the field of view of either the MOS or pn, and therefore do not have a spectrum from each camera.  After extraction, the spectra for each source are fit simultaneously using \sherpa, the data modeling and fitting package in CIAO 4.3 \citep{Freeman01, Fruscione06}.  We implement Nelder-Mead simplex optimization \citep{Nelder65} using Cstat statistics, a modification of the statistics of \citet{Cash79}.  For sources within the cluster half-light radius that have greater than 200 net counts in the EPIC cameras (X1, X2, X3, X4, and X5) we do model determination between five different XSPEC models: thermal bremsstrahlung, APEC thermal plasma \citep{Smith01}, blackbody, neutron star atmosphere \citep{Pavlov91, Zavlin96}, and power law.  All source models also take into account absorption, including an absorption component using the minimum neutral hydrogen column density of $N_{H}=8.3\times10^{20}$ cm$^{-2}$, fixed using the optical reddening value of NGC 6819 and is frozen for model fitting.  The APEC model abundances are fixed to solar to match the near solar metallicity of NGC 6819 \citep{Bragaglia01}.  Background spectra are extracted from source-free regions as near to the source as possible while attempting to keep a similar distance to the readout node, and are modeled simultaneously with the source using an absorbed power law.  

For the higher count sources the model with the best fit statistic and physically reasonable results is reported.  Goodness of fit is calculated using Cstat statistics, which is a suitable reduced statistic method for non-binned data.  We also report the Cstat statistic Q-value which is a measure of the probability that one would observe the given statistic value if the model is correct and the the best-fit parameters are true.  For lower count sources with less than 200 net counts, where model determination is more difficult, we fit only an absorbed power law with absorption in order to estimate the X-ray flux.  For sources within the half-light radius of the cluster we calculate the unabsorbed X-ray luminosity from the best-fit model using a distance of 2.3 kpc from recent \kepler\ data \citep{Basu11}.  We refer the reader to the XSPEC manual for more information on the implementation of Cstat \citep{Arnaud96}.

The \sherpa\ task {\task conf} is used to calculate confidence interval bounds, which varies a given model parameter value while allowing other free parameters to float to new best-fit values.  Uncertainties for model parameters are given as the {\task conf} $1\sigma$ confidence limits.  Flux uncertainties are obtained with the \sherpa\ task {\task sample\_energy\_flux}, in which the model parameter confidence intervals are randomly sampled 1000 times and the flux calculated for each realization.  The standard deviation of the resultant flux distribution is taken as the $1\sigma$ flux uncertainty.  Modeling results are listed in Tables~\ref{srctable1} and~\ref{srctable2} for the sources within and outside of the cluster half-light radius, respectively.  \\ \\

\section{X-RAY SOURCE CLASSIFICATION}
\label{sec:classification}

\begin{figure*}
\begin{center}
\includegraphics[scale=0.55]{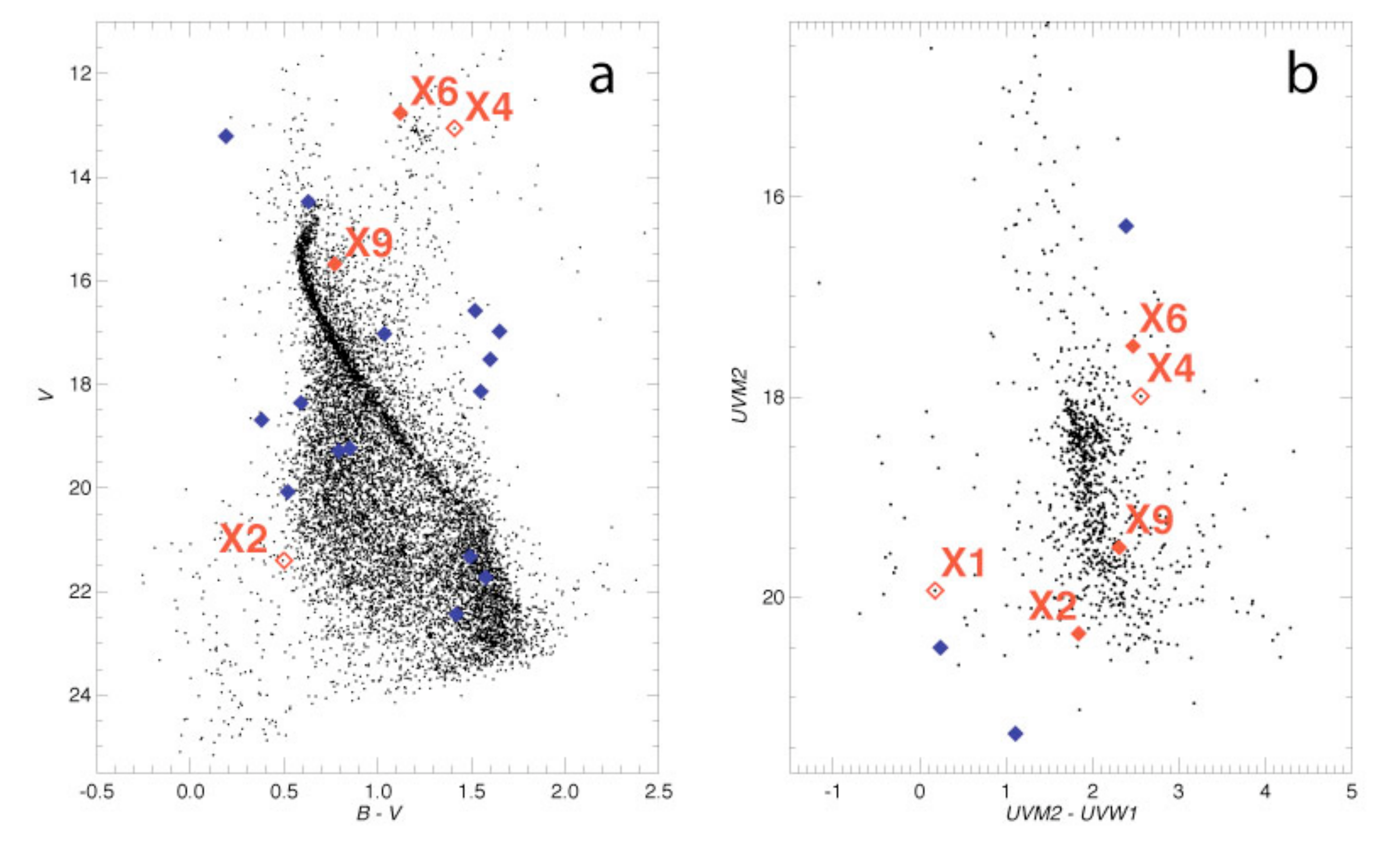}
\end{center}
\caption{NGC 6819 optical CMD (a) based on the photometry from \citet{Kalirai01b} and UV CMD (b) based on sources that are detected in both \textit{UVW1} and \textit{UVM2} filters.  Solid red diamonds correspond to X-ray sources within the diameter of NGC 6819 that have optical counterparts that fall within the X-ray position error.  Open red diamonds show possible optical counterparts that have a separation between $1\sigma$ and $1.5\sigma$ from the X-ray position.  Note that source X1 has a possible UV counterpart but does not have an optical counterpart.  Blue diamonds show X-ray counterparts for sources outside the diameter of NGC 6819, many of which have optical counterparts but no UV counterparts.}
\label{cmds}
\end{figure*}

Six of the 12 sources in NGC 6819 --- X5, X7, X8, X10, X11, and X12 --- are both faint and have no optical or UV counterparts.  We attempt to classify the other six sources --- X1, X2, X3, X4, X6, and X9 --- based on their X-ray and counterpart properties.  We note that source X3 lacks an optical or UV counterpart as well, but we still attempt to classify its emission because it is bright enough to carry out model determination.  Figures~\ref{cmds}a and b show the X-ray counterparts on optical and UV CMDs, respectively.  In both figures, solid diamonds correspond to probable counterparts whose distance from the X-ray source falls within the $1\sigma$ X-ray position error.  Open diamonds correspond to possible counterparts that have a separation between $1\sigma$ and $1.5\sigma$ from the X-ray position.

\begin{figure}
\begin{center}
 \includegraphics[scale=0.55]{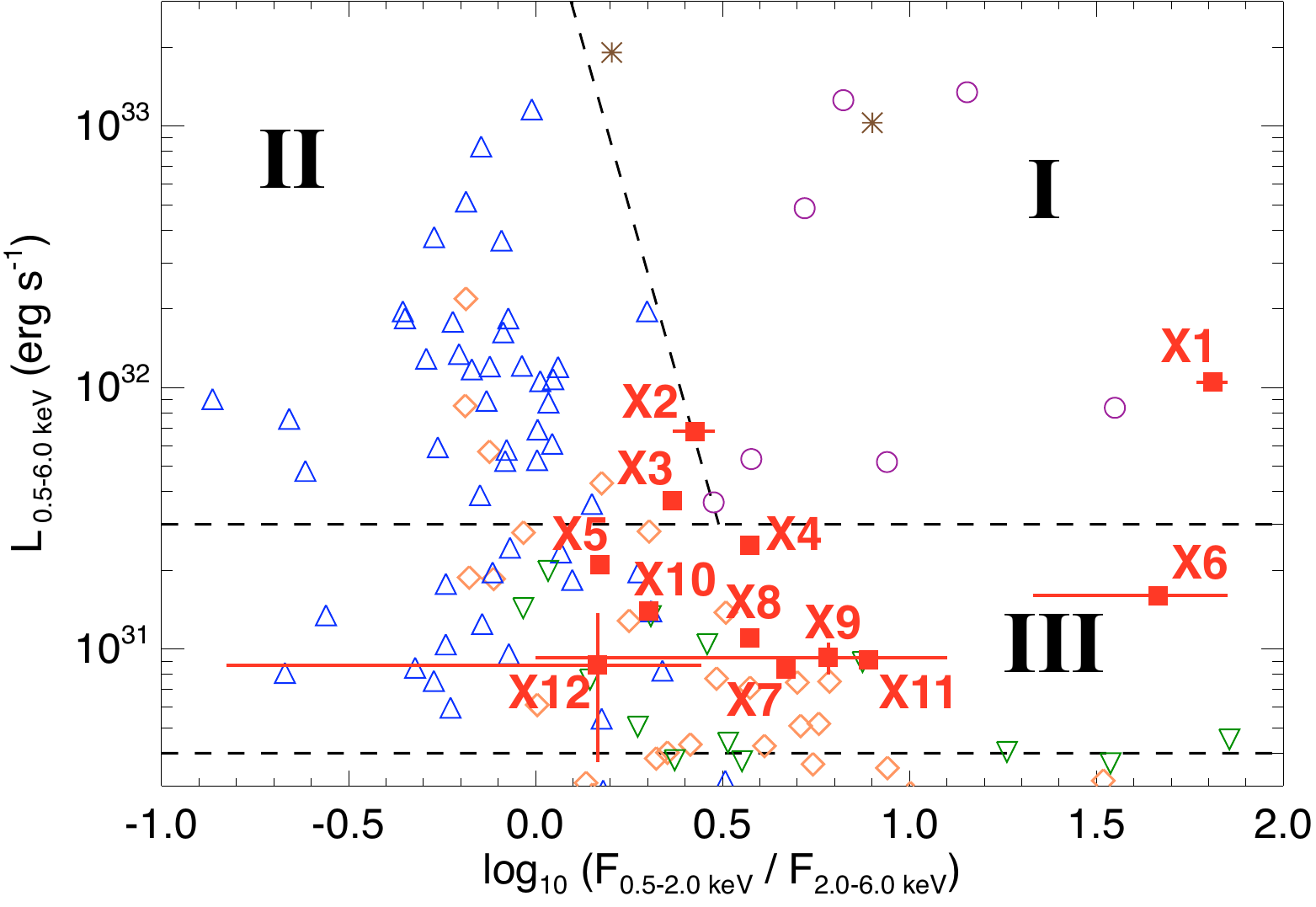}
\end{center}
\caption{X-ray CMD for sources within the half-light radius of NGC 6819, shown as filled red squares.  The x-axis shows the hardness ratio, or X-ray color, of the objects.  Soft counts are calculated from 0.2--2.0 keV and hard counts are calculated from 2.0--6.0 keV.  Securely identified globular cluster sources are shown as brown stars for LMXBs, purple circles for possible LMXBs, blue triangles for CVs, inverted green triangles for pulsars, and orange diamonds for active binaries \citep[see][and references therein]{Pooley06}.  From these identifications this diagram is roughly separated into three different X-ray populations of qLMXBs (population I), CVs (population II), and low-luminosity systems (population III) that has contributions from I and II  as well as most active binary systems.  The lines separating the populations are defined in \citet{Pooley06} for \chandra\ sources and are converted to \xmm\ values here.  Errors bars are shown in red for X1, X2, X6, X9, and X12, and are representative of other sources in the area.  The luminosity errors for X1, X2, and X6 are too small to be seen on this plot.}
\label{xcmd}
\end{figure}

Previous studies show that studying X-ray sources on a diagram comparing luminosity to X-ray hardness ratio is an effective diagnostic for source classification \citep{Pooley06}.  Hardness ratio here is defined as the number of soft counts (0.2--2.0 keV) divided by the number of hard counts (2.0--6.0 keV) and is essentially an X-ray color, so this diagram can be thought of as an X-ray color-magnitude diagram.  This type of diagnostic separates objects whose X-ray emission mechanisms have characteristically harder spectra, such as cataclysmic variables (CVs), or softer spectra, such as  quiescent low-mass X-ray binaries (qLMXBs).  In Figure~\ref{xcmd} we show an X-ray CMD for the 12 sources within the half-light radius of NGC 6819, shown as red squares.  We also show securely identified objects from globular cluster studies of various types as comparison \citep[see][and references therein]{Pooley06}.  Based on the extensive knowledge and confident identifications of X-ray sources in globular clusters, this diagram can be separated into population I, II, and III objects.  The division between population I and II objects is defined for \chandra\ sources in \citet{Pooley06} and has been converted to an \xmm\ color here.  Although there is some overlap between regions, population I objects are dominated by qLMXBs, population II objects are largely CVs but have some contribution from the most luminous active binaries, and population III objects are low-luminosity systems that most likely include contributions from both population I and II as well as millisecond pulsars (MSPs) and RS CVn or other active binary systems.

\begin{figure}
\begin{center}
\includegraphics[scale=0.5]{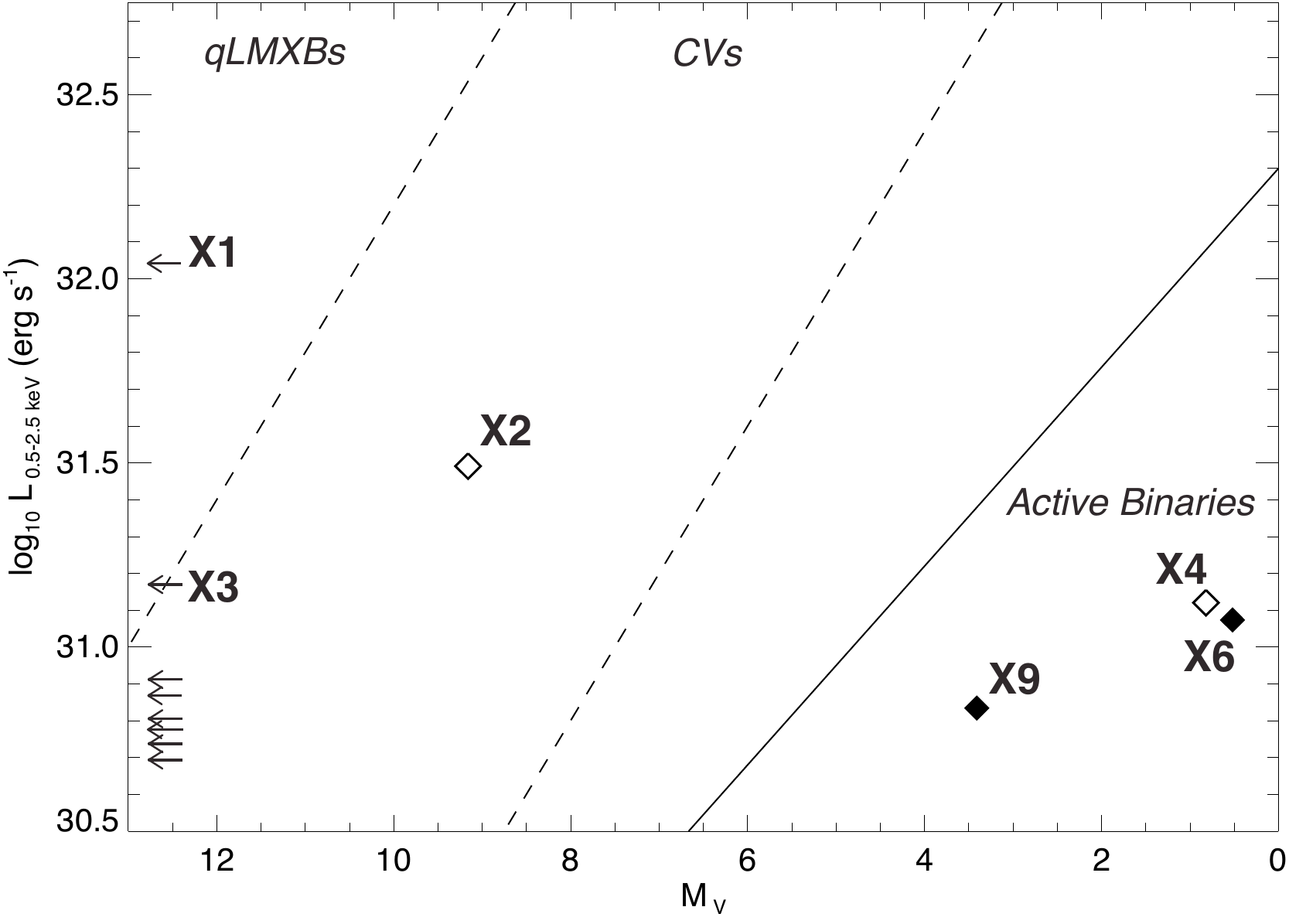}
\end{center}
\caption{X-ray to optical luminosity ratio for the X-ray sources within the half-light radius of NGC 6819, adapted from \citet{Verbunt08}.   Optical luminosity is given as the absolute $V$ magnitude of the optical counterpart and X-ray luminosity is given as the 0.5--2.5 keV luminosity.  Sources without optical counterparts are shown with arrows at the CFHT absolute magnitude detection limit of $V=12.76$ at the distance of NGC 6819.  The arrows in the bottom left, from top to bottom,  correspond to sources X8, X5, X11, X10, X7, and X12.  The dashed lines of $\log (L_{X}/L_{V}) = C$ separate various source types.  The top line ($C=36.2$) separates qLMXBs from CVs, and the bottom line ($C=34.0$) separates CVs from active binaries.  The solid line is an approximate upper limit to the X-ray to optical luminosity ratio for active binaries in the solar neighborhood.  Solid diamonds correspond to confident optical counterparts, while open diamonds correspond to possible counterparts that have a separation of $1\sigma$ to $1.5\sigma$ from the X-ray source.}
\label{xopt}
\end{figure}

Another diagnostic tool for determining the source of X-ray emission is comparing X-ray to optical luminosity ratios.  In Figure~\ref{xopt}, adapted from \citet{Verbunt08}, we show the ratio of X-ray luminosity to absolute $V$ magnitude for the NGC 6819 sources.  The optical absolute \textit{V} magnitudes correspond to the X-ray optical counterparts detailed in Table~\ref{match} or are shown as an arrow at the CFHT data absolute magnitude limit of $V=12.76$ at the distance of NGC 6819.  The X-ray luminosities for sources X1-X12 are given as the 0.5--2.5 keV luminosity.  The dashed lines separate different X-ray populations.  From \citet{Verbunt08} a line of constant X-ray to optical luminosity ratio given by $\log L_{0.5-2.5 \mathrm{keV}} (\textrm{erg s}^{-1}) = 36.2 - 0.4 M_{\mathrm{V}}$ separates qLMXBs (above the top line) from CVs, and a line given by $\log L_{0.5-2.5 \mathrm{keV}} (\textrm{erg s}^{-1}) = 34.0 - 0.4 M_{\mathrm{V}}$ separates CVs from magnetically active binaries (below the bottom line).  The solid line is an approximate upper bound to the X-ray to optical luminosity ratio for active binaries in the solar neighborhood given by $\log L_{0.5-2.5 \mathrm{keV}} (\textrm{erg s}^{-1}) = 32.3 - 0.27 M_{\mathrm{V}}$.  For more information on the basis of these separations please see \citet{Verbunt08} and references therein.

Through combining the information from these classification methods we conclude that X1 is a candidate qLMXB, X2 is a candidate CV, and X6 and X9 are RS CVn systems.  We are unable to classify X3 and X4 with these methods.

\subsection{Candidate qLMXB}
\begin{figure}
 \begin{center}
  \includegraphics[scale=0.6]{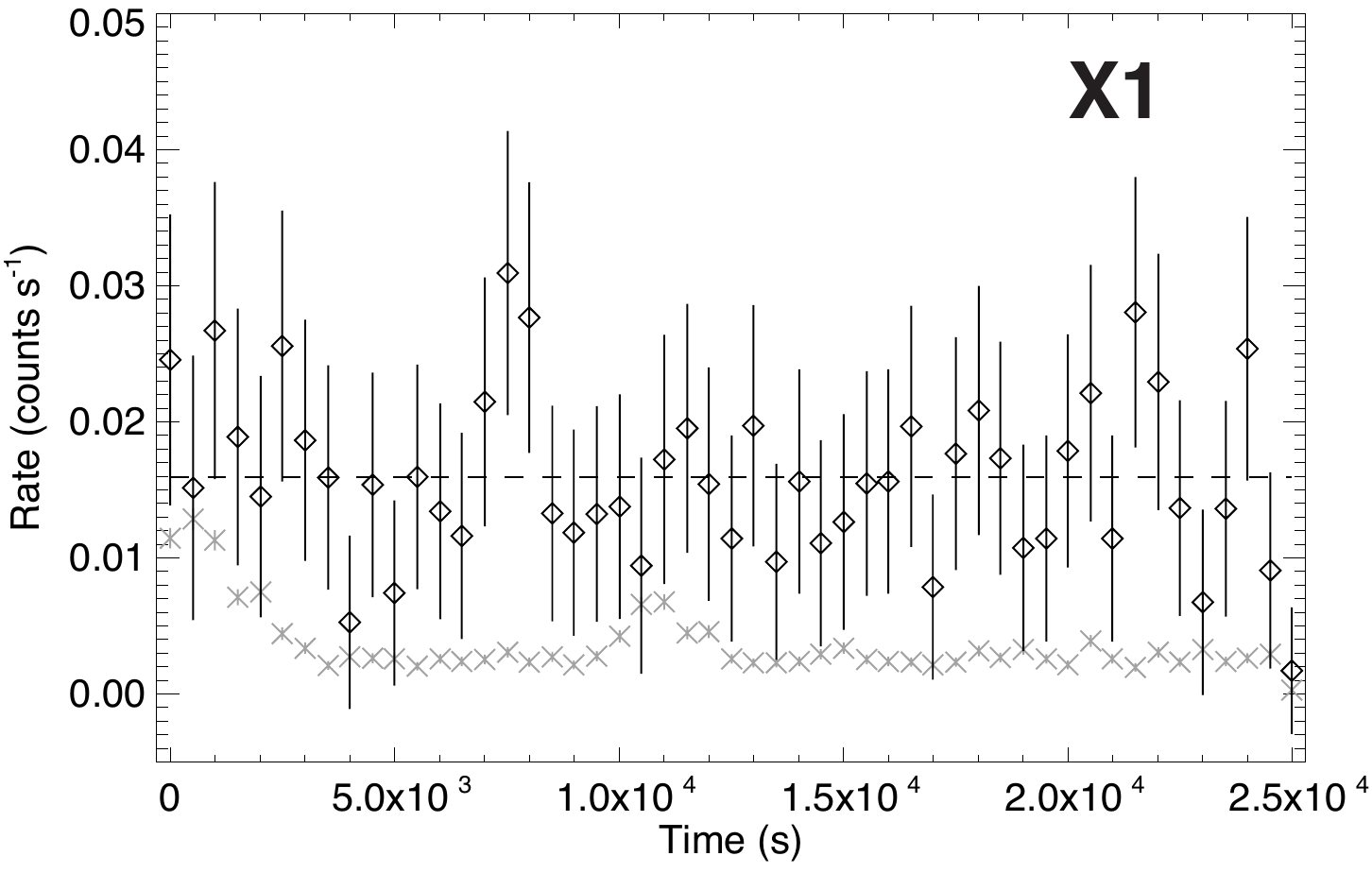}
 \end{center}
\caption{Background-subtracted light curve for X1 from the pn camera, with 500 second bins.  The errors are calculated using Gehrels approximation to confidence limits for a Poisson distribution \citep{Gehrels86}.  The appropriately scaled background light curve is shown in grey, and the average source rate of 0.016 counts s$^{-1}$ is shown with the dashed line. Statistical tests of the light curve against a constant model show X1 to be mostly non-variable except for a possible event at $\sim$7500\,s.}
\label{X1lc}
\end{figure}

In Figure~\ref{X1lc} we show a background-subtracted pn camera (2-10 keV) light curve for X1 (500s bins).  The appropriately scaled background light curve is shown in grey.  The light curve errors are calculated from the source counts using Gehrels approximation to confidence limits for a Poisson distribution \citep{Gehrels86}.  The average number of counts for X1 in the pn camera is 7.9 counts per bin, or a rate of 0.016 counts s$^{-1}$, and is shown with the dashed line.  Analyzing the resulting power spectrum yields no coherent periodicity in the source.  We also tested the light curve against a constant model, which yields a reduced $\chi^{2} = 1.2$ with 50 degrees of freedom ($p$-value = 0.16), indicating that X1 is consistent with being constant.  However, visual inspection of the light curve in Figure~\ref{X1lc} reveals a possible feature around $\sim$7500\,seconds.  That bin contains 15 counts; the probability of obtaining 15 counts per bin from a Poisson distribution with a mean of 7.9 counts per bin is only 1.6\%, indicating that the feature is likely due to stochastic variability.  Based on the ambiguous statistical evidence, we can say only that X1 is mostly constant except for a possible mild event in which the count rate roughly doubled for a period of $\sim$1000\,s.

The X-ray spectrum of source X1 is remarkably soft; almost all of its flux is below 2.0 keV, which strongly suggests that X1 is not a CV.  Regardless, we explore models typical of both CVs (APEC thermal plasma) and qLMXBs (neutron star atmosphere).  We find that no APEC model gives a reasonable fit to the X-ray spectrum.  The best fit was both statistically unacceptable with a reduced $\chi^{2}$ of 3.02 for a pn spectrum with 15 counts per bin {\it and} a temperature of only kT = 0.3 keV, which is well outside the range of typical CV temperatures \citep{Byckling10, Fertig11}.  

\begin{figure}
\begin{center}
\includegraphics[scale=0.55]{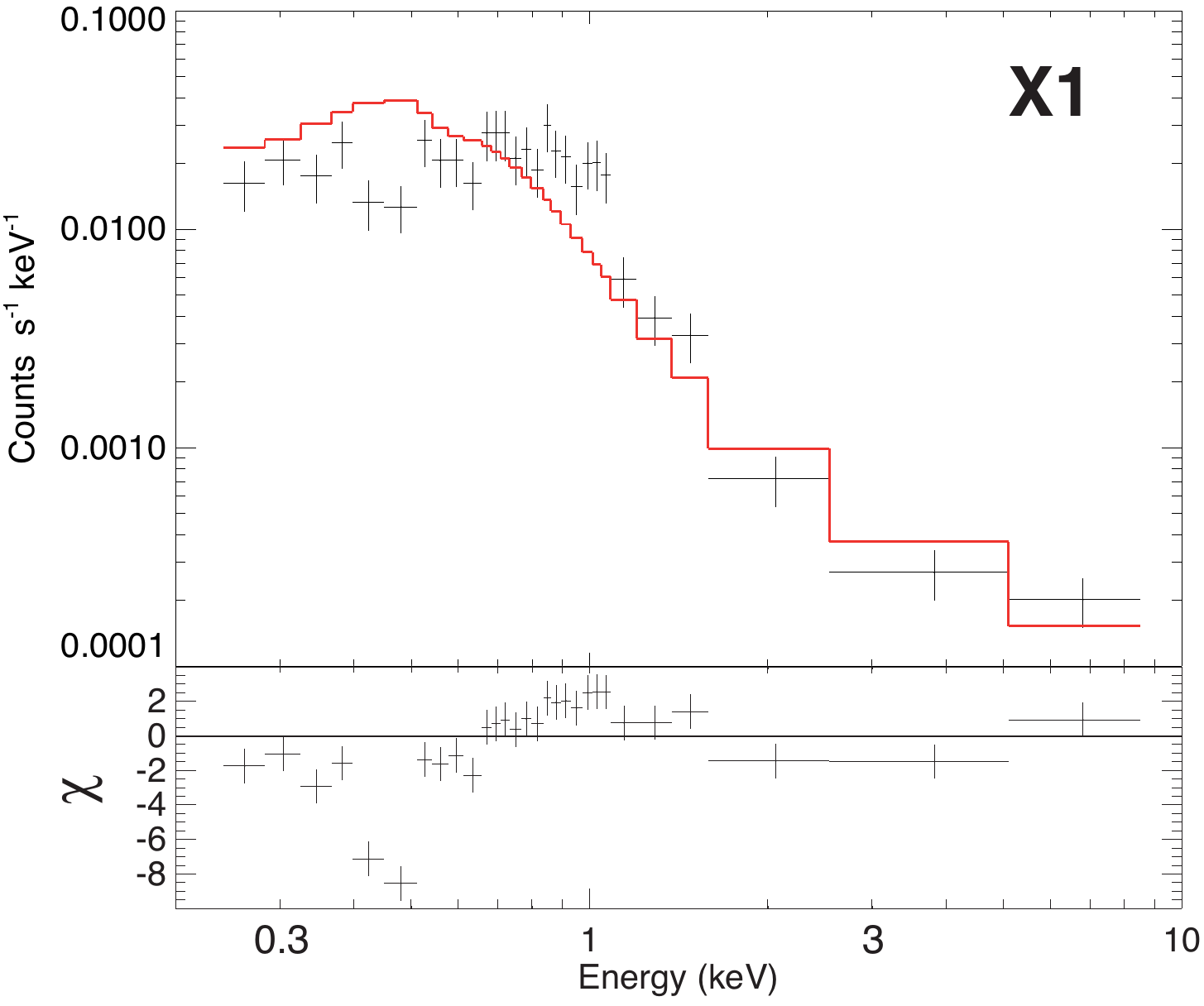}
\end{center}
\caption{Spectrum from the pn camera and best-fit neutron star atmosphere model for X1.  The data are binned for plotting purposes only with 15 counts per bin.  The errors are calculated using data variance $\chi^{2}$ statistics.  The residuals shown are the difference between the data and model divided by the uncertainty, plotted with error bars of unity to show the deviation in sigma between the model and the data.  We note that there is an unusual absorption component at $\sim$0.5 keV that is not explained by the neutron star atmosphere model.}
\label{X1spec}
\end{figure}

In contrast, a neutron star atmosphere fits quite well.  We fit the X-ray spectrum for X1 with a neutron star atmosphere model using the canonical neutron star values of a radius of 10 km and a mass of 1.4 M$_{\odot}$, setting the model normalization using the known distance to NGC 6819.  It is not necessary to include a hard power law component in the model to achieve a reasonable fit.  This is in agreement with the spectral properties of globular cluster qLMXBs which also lack a hard power law component, in contrast to field qLMXBs which often require an additional hard power law for successful spectral fitting \citep{Heinke03}.  Fitting X1 with a neutron star atmosphere model with an absorption component (as described in \S\ref{sec:xrayspec}) results in an effective temperature of 62.8$\pm0.5$ eV, which is within the typically accepted temperature values for qLMXB systems \citep{Heinke03}.  The pn spectrum and best-fit model with residuals are shown in Figure~\ref{X1spec}.  The data are binned for plotting purposes only and are shown with errors calculated using data variance $\chi^{2}$ statistics.  Residuals, calculated as the difference of the data and the model divided by the uncertainty, are plotted with error bars of unity to show the deviation in sigma between the model and the data.  We note there is an unusual absorption component at $\sim$0.5 keV, similar to what was seen in Aquila X-1 in quiescence by \citet{Rutledge02}.  Examination of the MOS spectra for X1 also show the same absorption.  The absorption feature cannot be explained by a neutron star atmosphere model alone, but perhaps may be indicative of metals present on the surface of the neutron star due to active accretion \citep{Rutledge02}.  However, further observations are necessary to determine whether the source of the absorption is inherent to the source and any possible physical significance it may have.

X1 is well separated from the other X-ray objects in Figure~\ref{xcmd}, lying within the area for population I, X-ray transient and qLMXB, objects.  There is a possible UV counterpart for X1 located 1\farcs45 (1.3$\sigma$) from the X-ray position.  The UV source is located in the region blue-ward of the main sequence in Figure~\ref{cmds}b, also known as the UV-excess region.  This region is primarily dominated by objects with accretion disk emission \citep{Dieball07}, and is a plausible location in a UV CMD for a qLMXB counterpart.  Although this is also the location in a UV CMD where CV populations are found, we note that the lack of coherent periodicity as well as the absence of a hard component to the X-ray spectrum makes it highly unlikely that X1 is an intermediate polar CV, so we tentatively classify X1 as a candidate qLMXB.  

The nearest optical source has $V=16.4$, $B - V=1.5$ and is located 1\farcs56 away from X1 (Figure~\ref{findingcharts}).  The UV counterpart and nearby optical source are separated by 0\farcs34, which is a $\sim$2$\sigma$ separation given the UV and optical source position errors, making the association of the two rather unlikely.  In addition, the optical source is far from the main sequence of NGC 6819, indicating that it is not a cluster member, and its red color is difficult to reconcile with UV emission.  The most plausible scenario is that the UV source is associated with X1 (separated by 1.3$\sigma$) and the optical source is an unrelated field star.  This nearby bright star makes detection of the true optical counterpart difficult, and it is likely that the true optical counterpart to X1 is hidden in the glare from this bright source.  Although we do not detect an optical counterpart, the X-ray to optical luminosity limit for X1 also shows it to be in the range of X-ray transient objects (Figure~\ref{xopt}).  We note that the lack of an optical counterpart is not surprising as qLMXBs can be extremely optically faint \citep{Heinke03}, with the added complication of a nearby bright source making detection of an optical counterpart very difficult.  

The X-ray spectrum, X-ray color, and limit of the X-ray to optical luminosity ratio all point to X1 as a candidate quiescent low mass X-ray binary.  In addition,  the absence of a hard component to the X-ray spectrum makes it highly unlikely that X1 is a dwarf nova or intermediate polar CV.  With this candidate qLMXB classification, X1 may be the first system of its kind found in an open cluster environment.  Higher spatial resolution X-ray and optical data are needed to confirm the counterpart identification and source classification.  We note that \citet{vandenBerg04} discovered a highly variable and soft X-ray source in M67 (CX 2).  They lack a confident classification due to the variable nature of the source, but detect system parameters that fall between expected values for a black hole qLMXB and a neutron star qLMXB.  However, \citeauthor{vandenBerg04} emphasize the need for more information before final classification.

\subsection{Candidate CV}
\begin{figure}
\begin{center}
\includegraphics[scale=0.55]{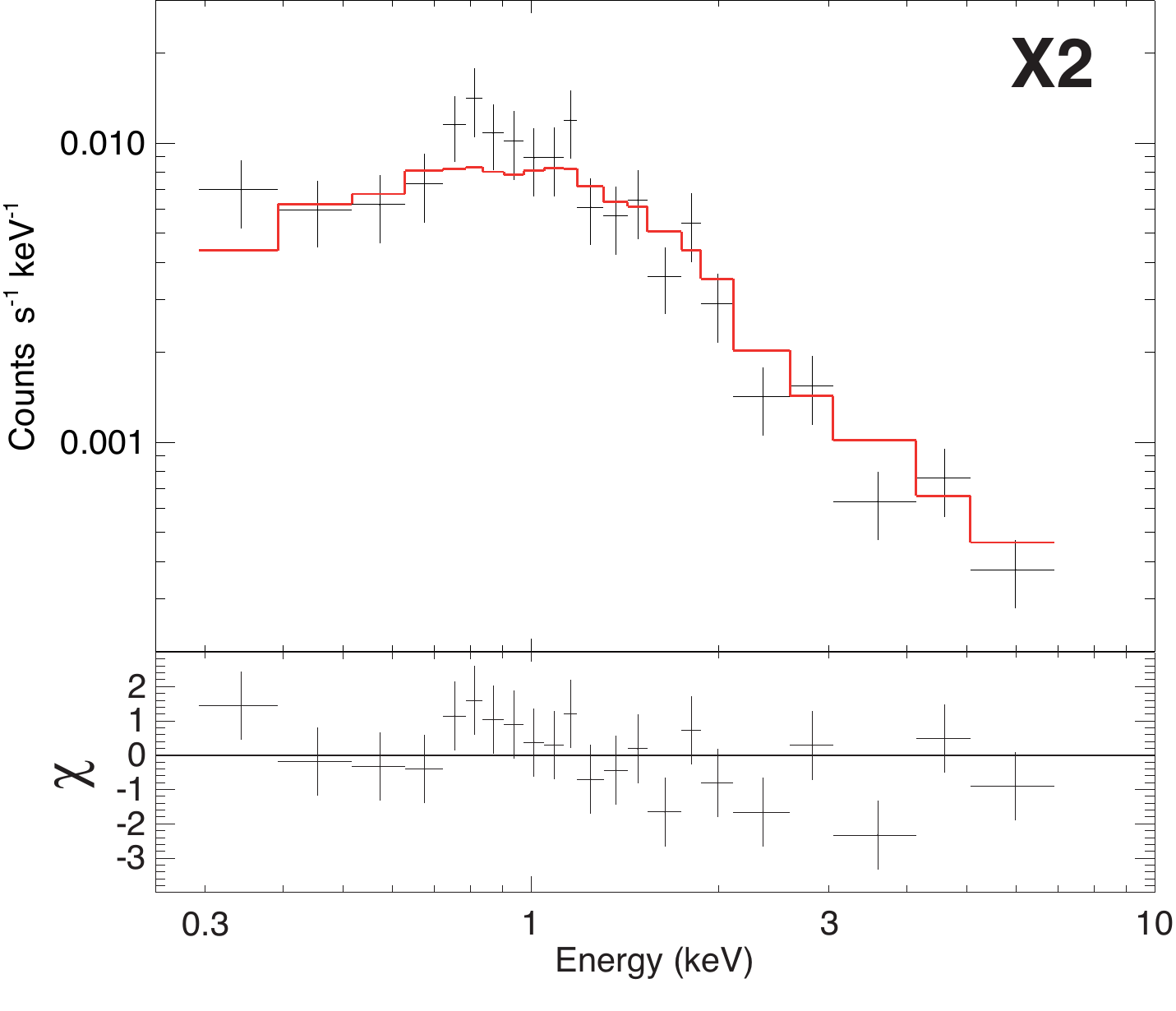}
\end{center}
\caption{Spectrum from the pn camera and best-fit APEC model for X2.  The data are binned to 15 counts per bin for plotting purposes only.   Errors are calculated using data variance $\chi^{2}$ statistics.  The residuals shown are the difference between the data and model divided by the uncertainty, plotted with error bars of unity to show the deviation in sigma between the model and the data.}
\label{X2spec}
\end{figure}

Source X2 has a confident UV source counterpart at 0\farcs79 (0.74$\sigma$) and a possible optical counterpart at a distance of 1\farcs24 ($1.15\sigma$).  The UV counterpart to X2 is located in the blue UV-excess region of Figure~\ref{cmds}b, an area dominated by objects with accretion disk emission such as CVs \citep{Dieball07}.  The possible optical counterpart is located in the ``gap'' region between the main-sequence and white dwarf cooling sequence (Figure~\ref{cmds}a), an area also dominated by CV systems.  Based on the X-ray color and luminosity of X2 (see Figure~\ref{xcmd}) it lies on the boundary between population I and II objects, indicating the possibility that this object is a CV candidate or qLMXB.  Its X-ray to optical luminosity ratio (Figure~\ref{xopt}) is more indicative of a CV than an LMXB, therefore we classify X2 as a CV candidate.  We fit the X-ray spectrum of X2 with an APEC thermal plasma model with an absorption component from the neutral hydrogen column density to the cluster, resulting in a best fit temperature of kT $=6.2^{+2.1}_{-1.0}$ keV, in good agreement with expected dwarf novae temperatures \citep{Byckling10, Fertig11}.  The pn spectrum and best-fit model with residuals for X2 is shown in Figure~\ref{X2spec}, binned to 15 counts per bin for plotting purposes only.  The error and residual calculations are the same as in Figure~\ref{X1spec}.

\begin{figure}
 \begin{center}
  \includegraphics[scale=0.6]{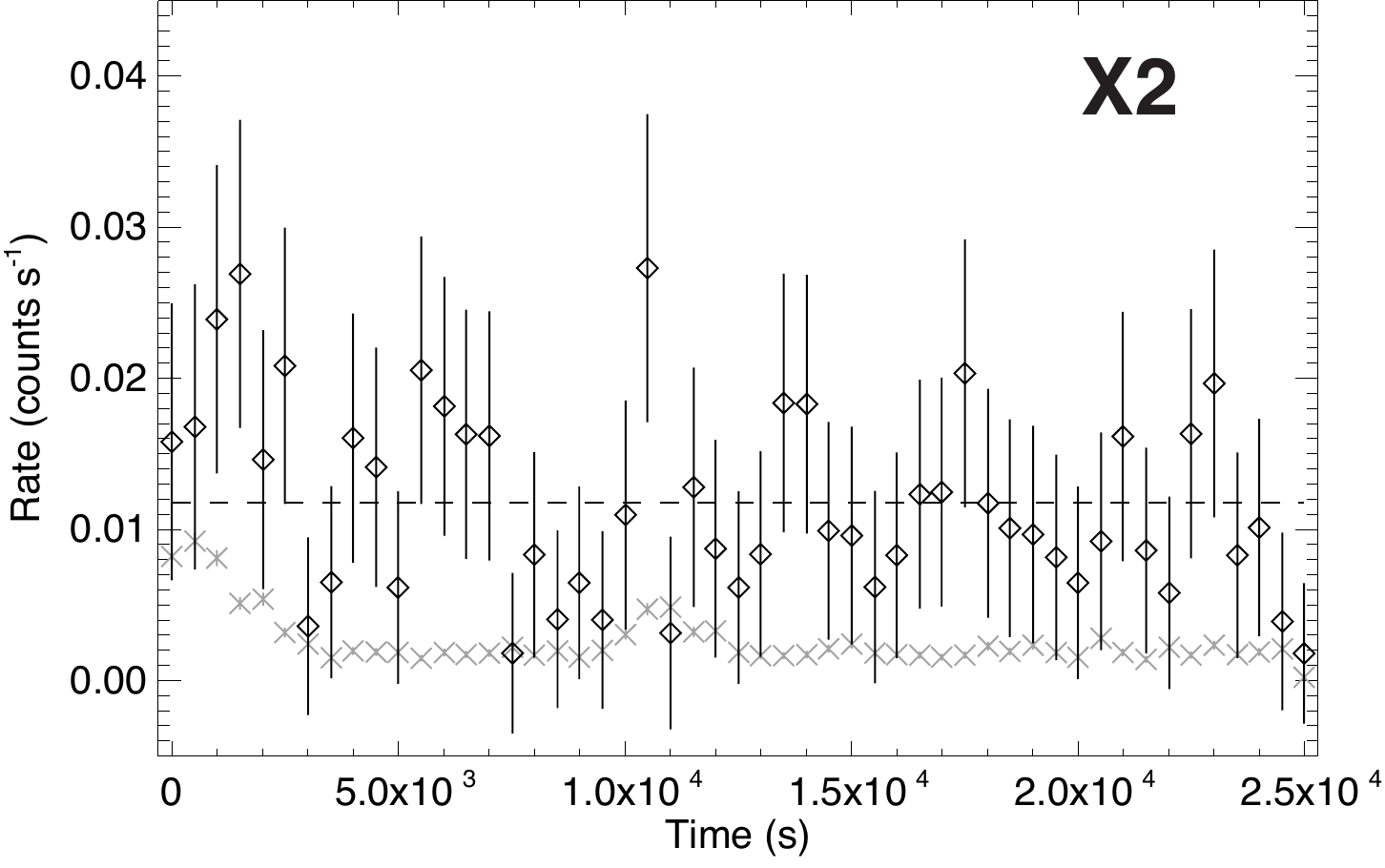}
 \end{center}
\caption{Background-subtracted light curve for X2 from the pn camera, with 500 second bins.  Errors are calculated using the same method as in Figure~\ref{X1lc}.  The appropriately scaled background light curve is shown in grey, and the average source rate of 0.012 counts s$^{-1}$ is shown with the dashed line.  A chi-square test of this light curve with 1000 second bins reveals X2 is not consistent with a constant source, although analyzing the power spectrum does not reveal any coherent periodicities.}
\label{X2lc}
\end{figure}

The background-subtracted light curve for X2, with 500s bins, is shown in Figure~\ref{X2lc}.  The appropriately scaled background light curve is shown in grey.  Errors are calculated using the same method as in Figure~\ref{X2lc}.  The average number of counts per bin is 5.8, or a rate of 0.012 counts s$^{-1}$, shown with the dashed line.  We find a bin with 13 counts, or a rate of 0.026 counts s$^{-1}$, as can be seen at $\sim$10500\,s.  The probability of finding a bin with 13 or more counts from a Poisson distribution with a mean of 5.8 is only 0.7\%.  A chi-square test of the light curve for X2 with 1000 second bins against a constant model yields a reduced $\chi^{2}$ = 1.8 with 50 degrees of freedom ($p$-value = 0.0005), which is inconsistent with a constant source.  However, analyzing the power spectrum for X2 does not reveal any coherent periodicity.  \citet{Eracleous91} show that CV X-ray spectra are often non-constant but lack a clear periodicity, so we take this light curve behavior to be consistent with the classification of X2 as a CV candidate.

\subsection{Active Binaries}
\label{sec:activebin}
X-ray source X6 has optical and UV counterparts at distances of 0.98 arcsec (0.7$\sigma$) and 1.16 arcsec (0.8$\sigma$).  Source X9 has optical and UV counterparts at distances of 1.32 arcsec (0.7$\sigma$) and 1.14 arcsec (0.6$\sigma$).  Both sources are confident cluster members due to radial-velocity measurements and/or proper motion membership probabilities (\citeauthor{Hole09} 2009; I. Platais, private communication).  Their X-ray to optical luminosity ratios are both indicative of active binary systems (Figure~\ref{xopt}), and their optical and UV color and luminosities are consistent with this classification (Figure~\ref{cmds}).

It is already known that the optical counterpart to X6, WOCS ID 3002, may have a dynamical history.  WOCS 3002 is a circular binary with a period of 17.7 days that appears in the red-clump of NGC 6819 (as seen in Figure~\ref{cmds}a).  If the primary is a horizontal branch star, the orbital separation of this system is too small to permit the primary star to exist at the tip of the red giant phase without significant mass transfer.  As such, \citet{Hole09} suggest that the system is the result of a dynamical encounter that exchanged the current primary star into the binary after its red giant stage, and studies have shown that this type of exchange is possible \citep{Gosnell07}.  Since it has an evolved primary star, we classify X6 as an RS CVn system.

Proper motion studies for the optical counterpart to X9 give a membership probability of 96\% (I. Platais, private communication).  The primary star in this system is a rapid rotator, which makes finding an orbital solution from the WOCS radial-velocity studies extremely difficult.  However, its X-ray to optical luminosity ratio and X-ray emission properties indicate an active binary system that we take to be a cluster member due to its proper motion.  On the UV CMD (Figure~\ref{cmds}b), source X9 is on the red edge of the region populated by main-sequence stars and red giants \citep[see Figure 4 in][]{Dieball07}.  X9 appears red of the main-sequence and is brighter than the equal-mass binary sequence on the optical CMD (Figure~\ref{cmds}a).  The optical location would seem to indicate a possible sub-subgiant binary system.  Sub-subgiants are known to exist in other open clusters as well, and have been shown to be a source of X-ray emission \citep{Mathieu03, vandenBerg04}.  We classify X9 as an active binary, noting that it is similar to an RS CVn but further observations are needed to verify the nature of the sub-subgiant system.  

\subsection{Unknown Objects}
Source X3 is difficult to classify.  The X-ray emission places X3 close to the intersection of all three population types in Figure~\ref{xcmd}, preventing a confident classification based on the X-ray information alone.  With no optical or UV counterpart classification becomes even more difficult.  The lack of any counterpart may indicate X3 is a background source, in which case our luminosity measurement is only a lower limit.  In addition, both blackbody and APEC thermal plasma spectral fits yield non-physical temperatures of 0.3 keV and 17 keV, respectively, when compared to the source hardness ratio.  Without more information we cannot determine the origin of the X-ray emission for X3 other than it is best fit with a power law index of $1.5^{+0.2}_{-0.8}$. 

The closest optical/UV counterpart for X4 is WOCS ID 4003, located at a distance of 1\farcs57 (1.3$\sigma$) from the X-ray position.  Radial-velocity studies of WOCS 4003 show it is not a short-period (P $< 3000$ days) binary, meaning it could be a single star or a long period binary.  WOCS 4003 has a low probability of being a cluster member from its radial-velocity, and proper-motion studies indicate 0\% probability that it is a member (\citet{Hole09}; I. Platais, private communication).  This object falls near the intersection of all three populations in Figure~\ref{xcmd} and in the area dominated by active binaries in Figure~\ref{xopt}, if WOCS 4003 is the true optical counterpart and assuming it is at the distance of NGC 6819.  Since cluster membership is highly unlikely, the luminosity given is almost certainly incorrect.  With only this information, the source of the X-ray emission for X4 is still unknown.

\section{DISCUSSION}
\label{sec:discussion}
Without a previous systematic study of the X-ray sources in many open clusters, it is difficult to predict the types and number of sources we expect to find.  If we extrapolate the relation between the number of X-ray sources and dynamical frequency in globular clusters to the regime of open clusters, we would expect almost no sources \citep[][see below]{Pooley06}.  Studies of two other individual open clusters have shown this not to be the case \citep{vandenBerg04,Gondoin05} but little is known about the formation of such populations in open cluster environments.  Here we make some reasonable assumptions about the formation of some of our sources, but definitive conclusions await accurate $N$-body modeling of NGC 6819 and systematic analyses of a sample of open cluster X-ray populations.

The presence of a qLMXB in an open cluster may be unexpected, and could be assumed to be dynamical given the known relationship between dynamics and qLMXB populations in globular clusters \citep{Pooley06}.  Although there are currently studies underway to determine the spatial density of qLMXBs in the field \citep{Grindlay05}, without such a measurement we are unable to determine how many primordial qLMXBs would be expected in an open cluster using field values.  \citet{Pooley06} investigate X-ray sources in globular clusters, fitting a two-component model to $n_{x}$ (number of X-ray sources per unit mass) vs. $\gamma$ (encounter frequency per unit mass) that separates the contribution from primordial and dynamical populations.  The model is applied to population I, II, and III objects separately as well as to all X-ray objects combined.  For population I objects (dominated by qLMXBs), the best fit to this relationship results in $0.4^{+0.53}_{-0.55}$ primordial sources per $10^{6}$ M$_{\odot}$.  Although this is a poorly constrained value, we apply it here to roughly determine the probability of finding a primordial population I object in NGC 6819.  \citet{Kalirai01b} integrate the mass function of NGC 6819 to find a cluster mass of $\sim$2600 M$_{\odot}$.  Using this value with the upper limit for primordial X-ray sources expected per unit mass, we expect $\sim$0.002 primordial population I objects in NGC 6819.  The low likelihood of finding a primordial qLMXB in NGC 6819 given this relationship would seem to indicate a dynamical history for X1.  

We apply this comparison with caution, noting that the validity of extrapolating globular cluster characteristics down to the regime of open clusters is dependent, in part, on whether the age difference between globular and open clusters affects the observable X-ray population.  \citet{Ivanova08} explore the number of qLMXBs in globular clusters over time, finding that the number of observable qLMXBs from 4 to 11 Gyr remains relatively constant for the majority of globular clusters, assuming that all qLMXBs that appear before 4 Gyr are primordial in nature.  This interesting result would seem to indicate that age does not play a major role in the number of observable qLMXBs, but the initial assumption that all sources prior to 4 Gyr are primordial limits the ability to confidently extrapolate their population characteristics back to the age of NGC 6819.  Regardless of the formation mechanism, the presence of a qLMXB is also unexpected given the results of \citet{Hurley05} that found with their $N$-body model only two neutron stars remaining in M67 at 4 Gyr, both of which are single stars.  Given the comparison tools currently available that all point to the unlikelihood of creating a primordial qLMXB in open cluster environment, we tentatively interpret X1 to be a dynamically formed candidate qLMXB.

Predicting the number of expected CVs in open clusters is similarly troublesome.  Although the CV populations of globular clusters and the field have been well studied, it is unknown whether either population is strictly applicable to rich open clusters at 2 Gyr.  We extrapolate the relationship for primordial CVs in globular clusters down to the masses of open clusters.  \citet{Pooley06} find a best fit of $0.27^{+1.40}_{-1.99}$ primordial population II objects (dominated by CVs) per $10^{6}$ M$_{\odot}$.  At the upper limit, this results in $\sim$0.004 primordial CVs in NGC 6819.  A more suitable comparison tool may be the field CV density.  The ChamPLane survey finds a local CV space density of $0.9^{+1.5}_{-0.5}\times10^{-5}$ pc$^{-3}$ at 95\% confidence, which is consistent with models of the galactic CV population \citep{Rogel08}.  This value is also consistent within two standard deviations to the local CV space density of \citet{Patterson98} at all scale heights.  Given the local mass density of the galactic plane of 0.1 M$_{\odot}$ pc$^{-3}$ \citep{Kuijken89}, this results in a CV mass density of $9.0^{+15}_{-5}\times10^{-5}$ CVs M$_{\odot}^{-1}$.  For the mass of NGC 6819 this results in a prediction of $0.23^{+0.39}_{-0.13}$ primordial CVs in the cluster.  The upper limit of this prediction, 0.62 CVs in NGC 6819, is not inconsistent with a detection of one CV in the cluster.  \citet{Hurley05} did create one primordial CV in M67 at a cluster mass of only 1400 M$_{\odot}$, which may point to primordial CVs being more prevalent in open clusters than either of these comparisons predict.  

The possible formation scenario for the optical counterpart of X6, outlined in \S\ref{sec:activebin}, may point to a dynamical history, but requires more observations and modeling to definitively determine the formation mechanism.  In addition, the possible sub-subgiant counterpart to X9 cannot be explained by single-star evolutionary theory.  Possible explanations require under- or over-luminous binary components that may be formed through mass transfer, stellar mergers, and/or dynamical stellar exchanges \citep{Mathieu03}.  Thus, while not yet definitive, several X-ray sources in NGC 6819 hint at dynamical formation processes.  The presence of dynamically formed X-ray sources in open clusters would indicate higher dynamical formation rates than seen in globular clusters.

Primordial X-ray sources in open clusters must also follow different trends than those seen in globular clusters.  Studies show that the X-ray sources of low density globular clusters scale more strongly with cluster mass than encounter frequency, which would indicate a population dominated by primordial sources \citep{Bassa08, Lu09, Lan10}.  The scaling factor for the number of primordial X-ray sources per unit mass, however, is low and predicts no close binary X-ray sources in the mass regime of rich open clusters.  It is possible that the low density environment of open clusters allows many more close binary products formed to persist, rather than be destroyed by the dense environment of globular cluster centers \citep{Davies97}.  We encourage further study into the theoretical modeling of creating close X-ray binaries in open cluster environments so that these formation pathways can be better understood. \\ \\ \\

\section{SUMMARY}
\label{sec:summary}
 
We conduct the first X-ray study of the rich open cluster NGC 6819, the first of eight open clusters in a survey of X-ray populations in open clusters using \xmmn.  Incorporating the use of Optical Monitor UV data and extensive optical photometry and spectroscopy from the CFHT Open Cluster Survey and WOCS, we implement a multi wavelength approach to classifying the X-ray population of NGC 6819.  We find 12 X-ray sources within the half-light radius of NGC 6819 to our luminosity limit of $10^{30}$ erg s$^{-1}$, where we only expect five or six background sources.  The X-ray sources include one of the first candidate qLMXBs in an open cluster environment (X1), a candidate CV (X2), and two active binary systems (X6 and X9).

The low probability of creating a primordial qLMXB in a cluster the size of NGC 6819 would seem to indicate that source X1 is most likely formed dynamically.  

The established relationship in \citet{Pooley06} between X-ray sources and encounter frequency in globular clusters predicts no X-ray sources in comparatively lower density open clusters.  Even if the sources are primordial, the known scaling of the number of primordial X-ray sources per unit mass in low-density globular clusters predicts no X-ray sources in the mass range of NGC 6819.  The detection of multiple X-ray sources in NGC 6819 indicates a departure from known relationships between X-ray sources and dynamics and mass in globular clusters.  It may be that primordial X-ray sources are able to remain in an open cluster environment in comparison to the higher density core of globular clusters \citep{Davies97}.  In addition, there may be unexpected dynamical creation processes responsible for the presence of X1, the candidate qLMXB.  We encourage further theoretical work in this area to help explain the creation of close X-ray binaries in lower density environments than has been previously investigated.  Completion of our X-ray study of open clusters is necessary in order to determine the full extent that dynamics play in creating X-ray sources in open cluster environments.

\acknowledgements We thank the anonymous referee for their helpful suggestions in improving this paper.  We gratefully acknowledge Imants Platais and thank him for sharing the proper motion data for NGC 6819 prior to publication.  N. M. G. would like to thank Thomas Nelson for his thoughtful comments in the production of this paper.  We acknowledge support from \xmm grant NNX08AY27G from the National Aeronautics and Space Administration.  N. M. G. and R. D. M. acknowledge National Science Foundation grant AST-0908082.

{\it Facilities:} \facility{XMM (EPIC, OM)}

\end{document}